\documentclass[11pt]{article}
 \usepackage{geometry}                % See geometry.pdf to learn the layout options. There are lots.
\usepackage{graphicx}

\usepackage{amsfonts, amsbsy, amssymb, amsmath, float,  color}
\usepackage{epstopdf}
\usepackage{multirow}
\usepackage{pdflscape}
\usepackage[utf8]{inputenc}
 \usepackage[footnotesize]{caption}

\begin{document}

\title{\textbf{Lagrangian study of the final warming in the southern stratosphere during 2002: Part I. The Vortex Splitting at Upper Levels}}

\author{J. Curbelo       \and
        C. R. Mechoso \and
        A. M. Mancho \and
        S. Wiggins} %etc.}}

\date{}

\maketitle

\begin{abstract}
The present two-part paper  provides a Lagrangian perspective of the final southern warming in 2002, during which the stratospheric polar vortex 
(SPV) 
experienced a unique splitting. Part I focuses on the understanding of fundamental processes for filamentation and ultimately for vortex splitting on a 
selected 
isentropic surface in the middle stratosphere.  Part II discusses the three-dimensional evolution of {the selected sudden warming event}. We approach the 
subject from a dynamical 
systems viewpoint and search for Lagrangian coherent structures using  a Lagrangian descriptor as a tool. In this Part I we work in the idealized framework of 
a kinematic model  that allows for an understanding of the contributing elements of   the flow in late September during the splitting. We  introduce a 
definition of kinematic 
SPV boundary based on a criterion for binning parcels inside and outside the vortex according to their values of the Lagrangian descriptor and associated PDF.  
This definition is justified by using arguments that go beyond heuristic considerations based on the potential vorticity.  Next, we turn to the filamentation 
proceses along this kinematic boundary, and the role of Lagrangian structures in the SPV splitting. We determine a criterion for splitting based on the 
structure of the evolving unstable and unstable manifolds.


\end{abstract}

\section{Introduction}
\label{intro}

We study the behaviour of the stratospheric polar vortex (SPV) in the southern hemisphere during the final warming in {spring (September-November)} of 
2002. This particular warming followed an evolution that was quite different from the norm as described by \cite{MOPF88} and {reviewed} by \cite{char}.
The SPV experienced a unique splitting at upper levels in late September, after which it reformed with a weaker strength to finally transition into the summer 
circulation. Several papers present detailed  {\it Eulerian} diagnoses of this final warming and vortex splitting 
\cite{Varotsos02,Varotsos03,Varotsos04,Allen2003,char,Konopka2005,Orsolini2005,Esler2005,man06,Taguchi2014}. 
The present two-part paper gives a {\it Lagrangian} perspective  of the event.

In this Part I we concentrate on a single potential temperature level in the middle stratosphere.  The analysis is carried out in the idealized framework of a 
two-dimensional (2D) kinematic model (KM).  The KM aims at emulating the behavior of the longest planetary waves on an isentropic surface as obtained from the 
reanalysis data, and allows for a detailed parametric study that is greatly facilitated by the dynamical systems approach. According to our previous work 
\cite{kinm}, a  2D KM with an appropriate choice of functional and geometrical features and model parameters can capture the fundamental mechanisms 
responsible for the complex fluid parcel evolution during polar vortex filamentation and breakdown. The work in the KM framework has two main objectives.  
The 
first is to explore a kinematic definition for the SPV boundary that can be justified, at least {partially}, with arguments that go beyond heuristic 
considerations based on the potential vorticity (PV) field as is generally done.  Such a definition will be used in Part II for the selection of parcel 
trajectories that can help visualize the 3D evolution of the event. The second objective is to determine a criterion for vortex splitting based on the evolving 
Lagrangian structures. Such a criterion will help in Part II to understand why the vortex splitting only occurred at some levels.

\begin{figure*}
\begin{tabular}{ll}
(a)  9 September 2002 & (b) 22 September 2002\\
\includegraphics[scale=1.1]{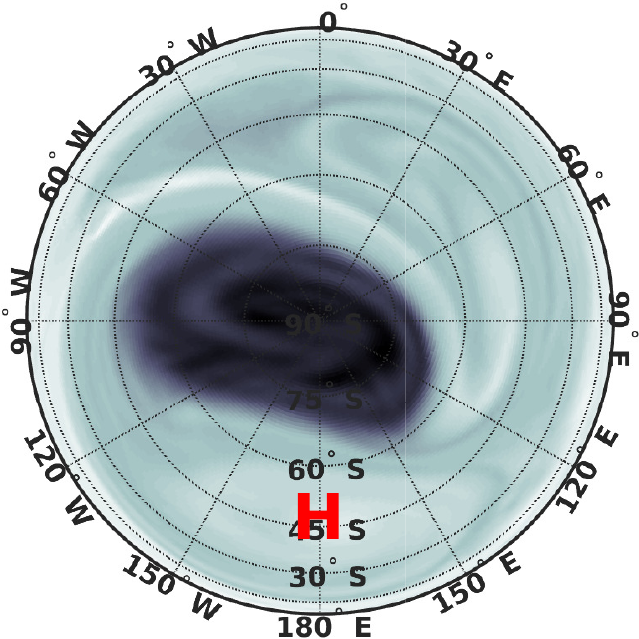}&\includegraphics[scale=1.1]{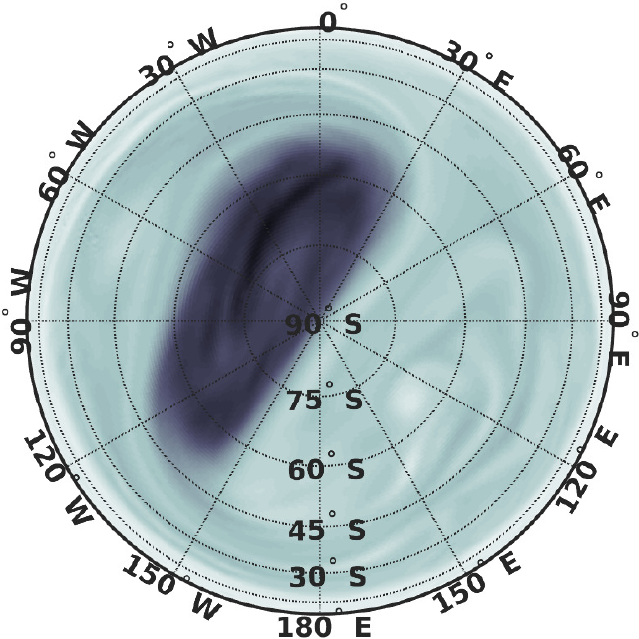}\\
(c) 24 September 2002 & (d)  15 October 2002\\
\includegraphics[scale=1.1]{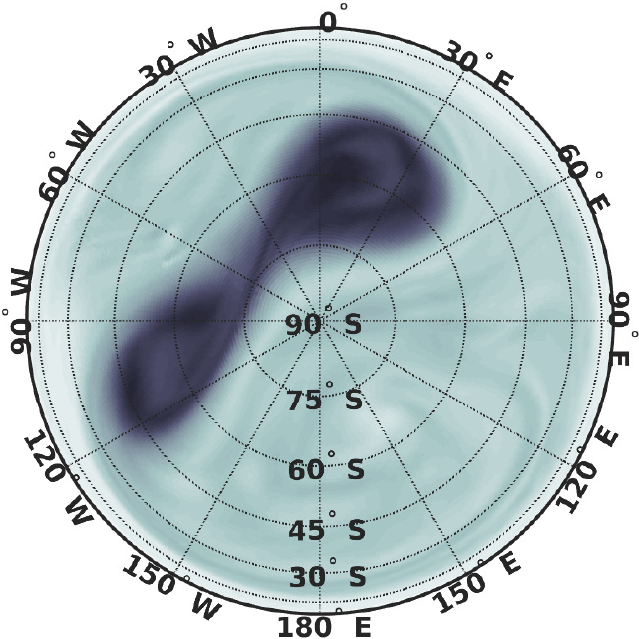}&\includegraphics[scale=1.1]{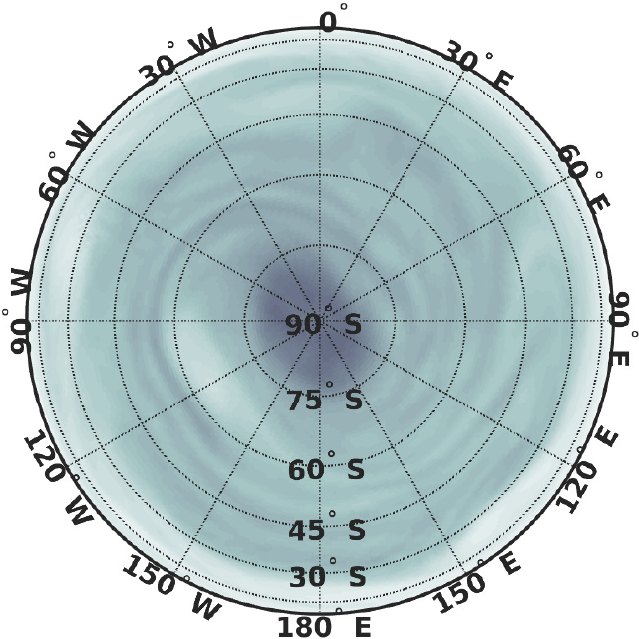}\\
\end{tabular}
\centering 
\includegraphics[scale=0.45]{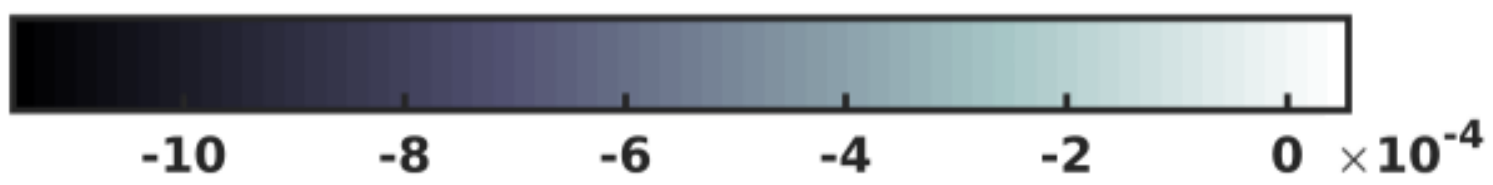}
\caption{Maps of potential vorticity at 850K for distinct stages of the southern final warming in 2002. Units are K m$^2$kg$^{-1}$s$^{-1}$. {The red 
letter 'H' in (a) indicates the anticyclonic center over the Southern Pacific}} 
\label{fig:PV}
\end{figure*}

We start in section \ref{s:2002} with snapshots of {\it Eulerian} features of the flow during the selected final warming event using a reanalysis dataset. 
Section \ref{s:methodology} outlines our methodology presenting the Lagrangian descriptor and describing the KM. Section \ref{s:lagrangianstructures} 
introduces 
a definition for the vortex boundary based on the Lagrangian descriptor. Section \ref{s:criterion} 
discusses a criterion for SPV split based on the evolving Lagrangian structures. Finally, section \ref{s:summary} presents a summary and the conclusions. Annex 
\ref{A:concepts} is a succinct review of Lagrangian concepts used in the paper.

\section{The southern final warming in Spring 2002}
\label{s:2002}

In this section we compile a selection of outstanding {\it Eulerian} features of the flow observed during the southern stratospheric final warming event of 
2002.  All figures in the present paper that refer to observed features are obtained using the ERA-Interim reanalysis dataset produced by the European Centre 
for Medium-Range Weather Forecasts (ECMWF; \cite{simmons07}). Figure \ref{fig:PV} displays PV maps at 850K in the middle stratosphere at distinct stages of the 
final warming. PV is given by,

\begin{equation}
PV= -g (f + \zeta_\theta) \frac{\partial \theta}{\partial p}
\end{equation} 

\noindent
where $f$ is the Coriolis parameter, $\zeta_\theta$ is relative vorticity on the isentropic surface $\theta$, and $ \frac{\partial \theta}{\partial p}$ is a 
measure of static stability. This expression includes only the vertical component of the vorticity and  uses the hydrostatic assumption. At the 850K level, the 
SPV in the early part of the period (September 9) was in a circumpolar configuration, albeit its center was slightly displaced from the South Pole, while an 
anticyclone had developed over the southern Pacific Ocean. A dramatic series of events ensued, during which the cyclonic vortex strongly elongated and split 
into two (September 22 and 24, respectively).  The two cyclonic vortices resulting from the splitting evolved further in time while interacting with each 
other, 
until one of them set in a polar position and the other disappeared as a separate entity.  By 15 October a weak cyclonic circulation remained around the pole 
without a robust structure at 850K.

\begin{figure*}
\begin{tabular}{ll}
(a) \\ \includegraphics[scale=0.545]{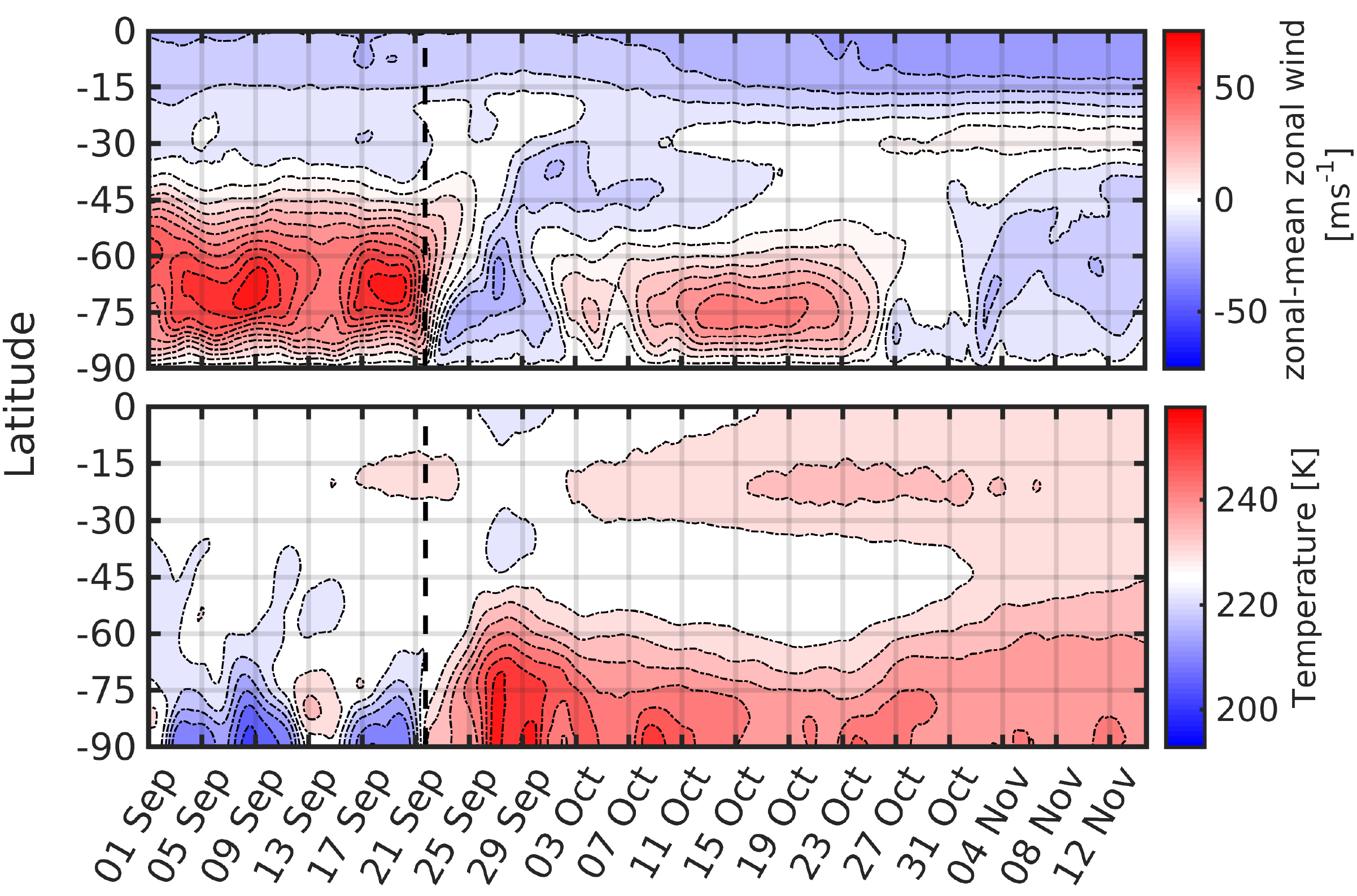}\\ 
(b) \\ \includegraphics[scale=0.5]{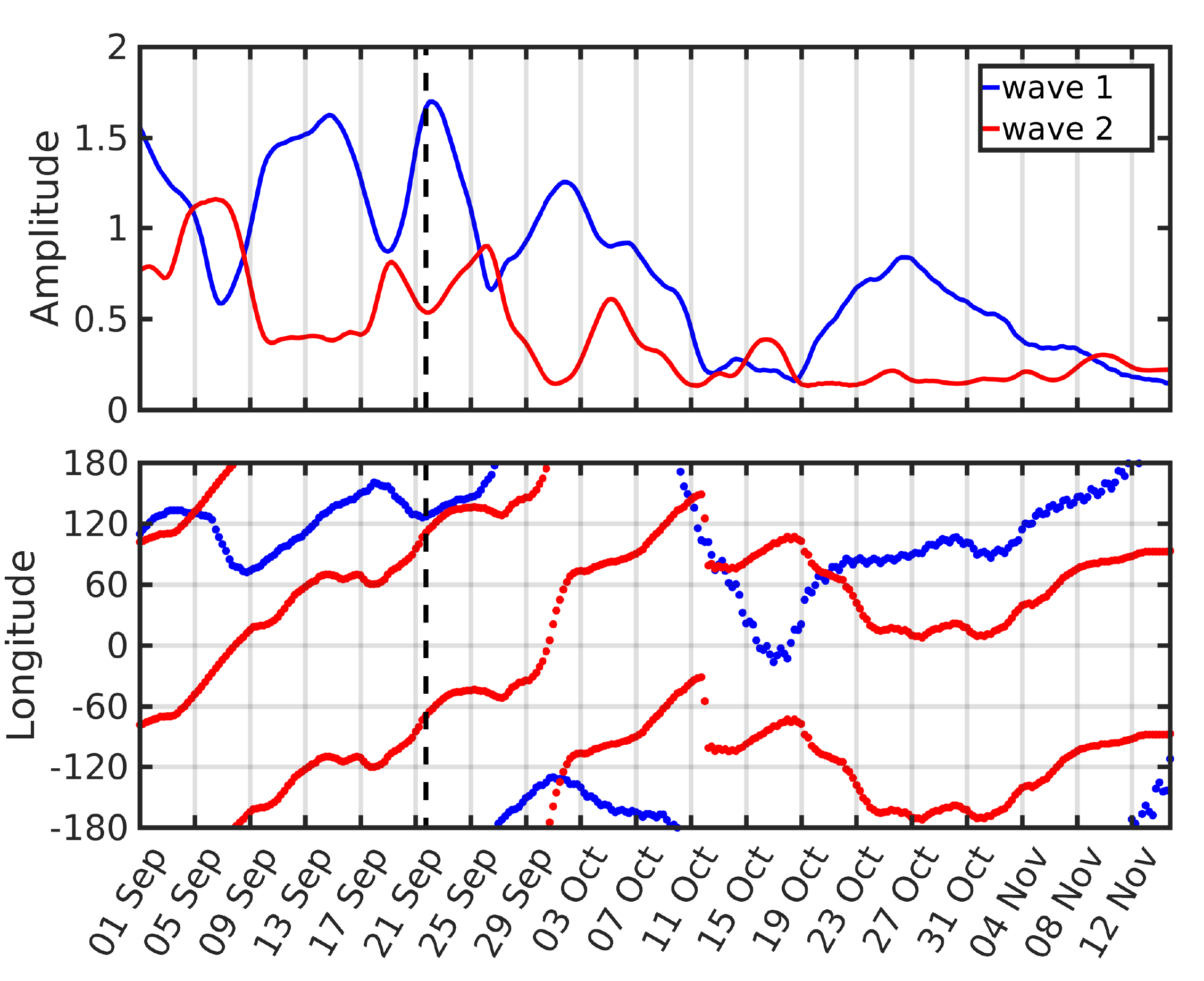}\\
\end{tabular}
\caption{{\bf Eulerian} diagnostics {at 10 hPa}:  (a) Time - latitude diagrams at 10 hPa of the zonal mean velocity (m s$^{-1}$) and 
temperature (K). (b) Time 
series at $60^\circ$ S for wave 1 (blue) and wave 2 (red), amplitudes {[km]} {(upper panel)} and location of the ridges {(lower panel)}. {September 22 
is marked with a vertical black dashed line}. }
\label{fig:Euleriandiagnostics} 
\end{figure*}

A complementary view of the event is provided in Figure \ref{fig:Euleriandiagnostics} for the 10 hPa level ($\sim$ 850K). {The upper part of panel (a)} in 
this figure 
represents the zonal velocity averaged along latitudinal lines ({zonal-mean zonal wind}), showing {strong westerlies} at high southern latitudes 
{until 2 days before} an 
abrupt change in direction {to easterlies} around September 22. {The lower part of} panel (a) displays the sharp increase of 
zonal mean temperatures in the 
polar region associated to this event. These changes in velocities and  {temperature} represent the only documented sudden stratospheric warming in the 
Southern 
Hemisphere on record. Panel (b) shows that planetary waves in the southern stratosphere were very active during the southern spring of 2002, 
particularly in September.  {Within} the ten-day period centered on September 22,  {the upper part of panel (b)} shows that the amplitude of wave 1 almost 
doubled, 
while the 
amplitude of 
wave 2 decreased, albeit not so drastically. {The lower part of} panel (b) confirms that wave 1 was quasi-stationary and wave 2 propagated eastward, as is 
typical in 
the 
southern stratosphere during early spring {\cite{mechoso1982,Bowman90,Manney91}}. 
{The period of wave 2 is around 19 days which gives a zonal angular phase speed of approximately  9.5$^\circ$/day.}
We note that increased amplitude of wave 1 results in 
a 
displacement of the SPV 
from a circumpolar configuration, while increased amplitude of wave 2 results in a stretching of the SPV in one  direction and contraction (or ``pinching'') in 
the orthogonal direction.

\section{Methodology}
\label{s:methodology}
\subsection{The Lagrangian descriptor  function M}
\label{ss:M}

As in our previous studies on the southern SPV(\cite{alvaro2, kinm}, \cite{JC17}) we follow an approach based on dynamical systems theory by searching for  
dynamical objects from which the general behaviour of particles can be deduced. Among these objects are  hyperbolic trajectories and their  invariant manifolds 
as well as others defined in the Annex.

Our principal analysis tool {for stratospheric studies} is the Lagrangian descriptor known as the function $M$. In the following paragraph we define this tool 
and give our motivation for its selection. The function $M$ is given by the expression, 

\begin{equation}
M(\mathbf{x}_{0},t_0,\tau) = \int_{t_0-\tau}^{t_0+\tau}\|\mathbf{v}(\mathbf{x}(t;\mathbf{x}_0),t)\| \; dt \;,
\label{M} 
\end{equation} 

\noindent
where $\mathbf{v}(\mathbf{x},t)$ is the velocity field and $\| \cdot\|$ denotes the Euclidean norm. At a given time $t_0$, $M$ corresponds to the length of the 
trajectory of a fluid parcel that starts at $x_0$ and evolves forwards and backwards in time for a time interval $\tau$. 
\begin{figure*}
\begin{tabular}{ll}
(a) 9 September 2002 & (b) 22 September 2002\\
\includegraphics[scale=1.1]{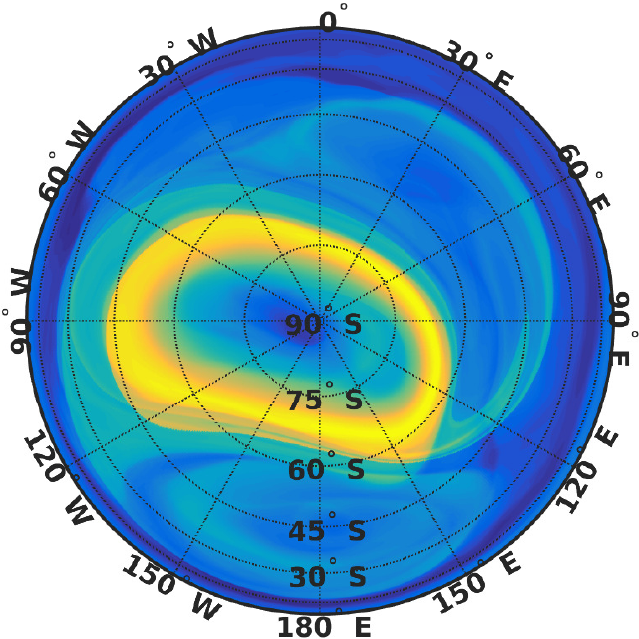}& 
\includegraphics[scale=1.1]{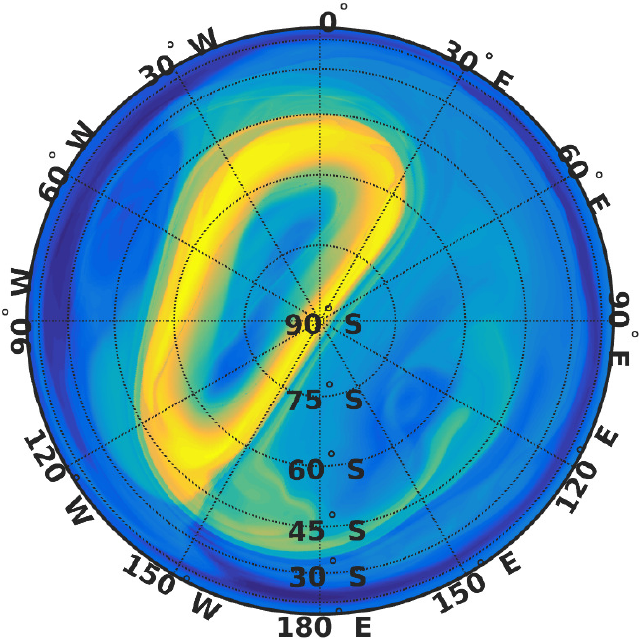}\\
(c) 24 September 2002 & (d) 15 October 2002\\
\includegraphics[scale=1.1]{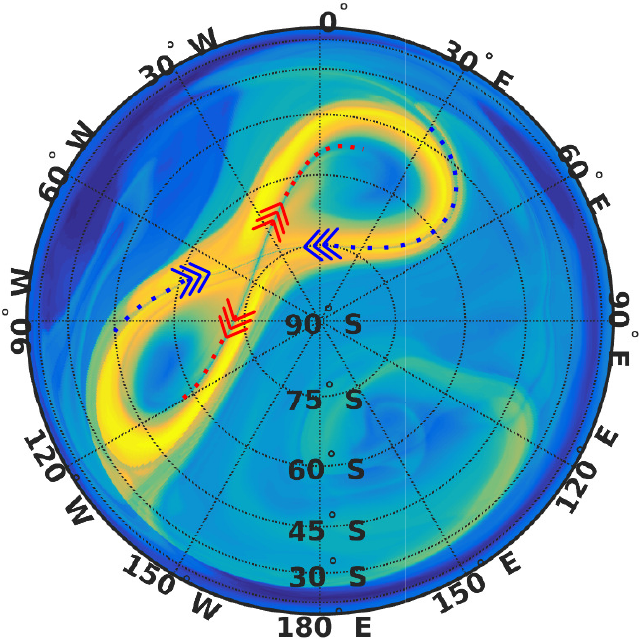}&\includegraphics[scale=1.1]{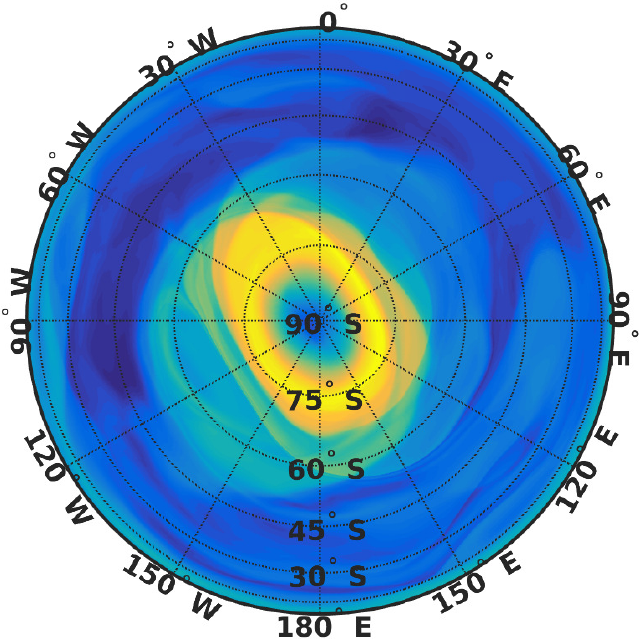}
\end{tabular}
\centering
\includegraphics[scale=0.45]{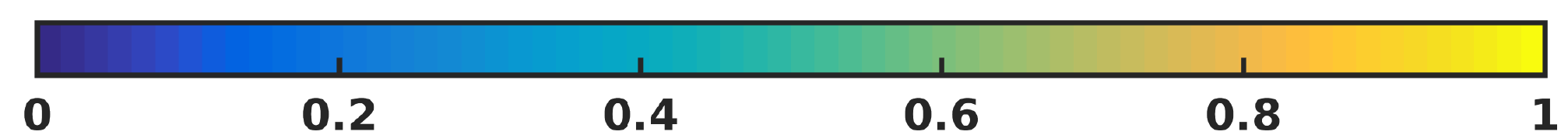}
\caption{The normalized function  $M$ ($\tau=5$) at 850K for the same days as in Figure \ref{fig:PV}. In all figures of $M$ in this paper, the bright yellow 
color represents  high values (i.e., long parcel trajectories  {in $(t_0 - \tau, t_0 + \tau)$}) and the dark blue color represents small values (i.e., short 
parcel trajectories  {in $(t_0 - \tau, t_0 + \tau)$}). {The blue and red arrows in (c) indicate the stable and unstable manifolds respectively. }}
\label{fig:M}
\end{figure*}

Despite its simple physical interpretation, the function $M$ is a powerful tool for finding the dynamical objects of interest for our study 
\cite{mm2009,mm2010,mwcm2013,ld_ijbc}. Stratospheric flows on the timescale of sudden warmings ($\sim$ 10 days) are, to a first approximation, adiabatic and 
frictionless, which justifies their study in 2D isentropic (i.e. constant potential temperature) surfaces.  Systematic numerical experimentation by 
\cite{mm2010,mm2012} have demonstrated that singular features of $M$ in 2D flows are aligned with invariant curves of the system (invariant manifolds). 
Moreover, the crossing points of such invariant curves represent trajectories of system. On the basis of such findings, \cite{mm2010,mm2012} conjectured that 
for sufficiently large values of $\tau$, large magnitudes of $\nabla M$ highlight the stable and unstable manifolds and, at  their crossings, hyperbolic 
points. 
Some analytical results in this regard are discussed in \cite{mwcm2013}. Accordingly, cross-hatched patterns of $M$ in a color-coded plot can be associated 
with 
the intersection of unstable and stable manifolds and associated hyperbolic points, whose instantaneous locations for time-dependent flows are points on 
hyperbolic trajectories.  \cite{ld_ijbc} discuss particular examples in which contour plots of a Lagrangian descriptor may not clearly capture the presence of 
singularities, but the velocity fields in these examples are not representative of the ones found in the geophysical context considered in the present paper. 
In 
view of these demonstrated properties, $M$ has already been successfully  applied in studies on dynamical and transport processes in the stratosphere 
(\cite{alvaro1, alvaro2,nz,anirban,manney2016,kinm}).

Figure \ref{fig:M} shows stereographic projections of the function $M$ evaluated using $\tau = 5$ days on the 850 K isentropic surface for the same days as 
Figure \ref{fig:PV}. {In this case, the 
trajectories are calculated using the horizontal wind velocity components (interpolated to isentropic surfaces)  which are 
available from the ERA-Interim four times daily (00:00, 06:00, 12:00, 18:00 UTC) with a horizontal resolution of $0.75^o \times 0.75^o$ in longitude and 
latitude. Details on the calculation of trajectories from a given velocity field can be found in Garcia-Garrido et al, (2017).} The pattern similarity between 
PV and $M$ plots is strong.  The PV plots emphasize the high values associated with the SPV and the detailed structure of the surrounding flow in what is known 
as the ``surf zone'' where Rossby waves break (see \cite{mcintyre84}). 
High (low) values  {of $M$}  correspond to long (short) trajectories  {during the time interval $(t_0-\tau, t_0+\tau)$, as is the case }  at locations along 
(away from) the jets. The time averaging in $M$ can blur some features with small spatial and time scales, but the ability to provide the locations of the 
hyperbolic trajectories has many advantages for stratospheric studies. Thus, the PV and $M$ diagnostics complement each other well for Lagrangian analyses of 
the stratospheric circulation.

\subsection{The kinematic model (KM)}
\label{ss:KM}

{The flow in the real atmosphere is very complex.  To gain insight into the fundamental mechanisms at work for transport processes it is useful to have a 
simple 
model that focuses on the behavior of the longest planetary waves and that allows for a series of sensitivity experiments. Therefore, we have  devised a simple 
KM in which the amplitude and phase speed of the waves can be specified analytically or according to observations.} \cite{kinm} demonstrate that one such model 
with a suitable choice of parameters is able to produce strikingly similar transport features to those found in the reanalysis data, including filamentation 
and 
vortex breakdown.  The KM in  \cite{kinm} 
consists of a vortex centered at the origin of a cylindrical coordinate system $(r, \lambda)$ plus two perturbations representing the Fourier components with 
wavenumbers 1 and 2 along ``latitude circles'' (lines of constant $r$); these perturbations can rotate in time around the origin.  The vortex and perturbations 
dependence on $r$ are chosen to mimic the structures of the zonal mean jet and longest planetary waves in September 2002.  
\cite{kinm} gives details on the KM numerics and calculation of trajectories.

In the present paper we use a slightly modified version of the KM in \cite{kinm}.  Let us consider the cylindrical system of coordinates $(r,\lambda)$ centered 
at $(0,0)$. In our KM, the velocity field corresponds to an axisymmetric vortex centered at 
$(x_0, y_0)=(r_0 \cos(\lambda_0), r_0\sin(\lambda_0))$, plus a perturbation with wavenumber 2 only around $(0,0)$. Thus, in our case, the effect of  wave 1 
(and 
other harmonics) appears in the displacement of the vortex from the origin $(0,0)$.  This modification allows for non-zero velocities across the origin 
$(0,0)$, 
which are not defined in the version by \cite{kinm}, but were present during the vortex splitting under consideration. The equation for the streamfunction in 
our KM is given by,

\begin{equation}
\Psi(r', r, \lambda, t) = \varepsilon_0(t) F_0(r')+ \varepsilon_2(t) \Psi_2(r,\lambda),\label{eqpsi}
\end{equation}
where $r'$ is the distance to $(x_0, y_0)$.  The first term in the right-hand-side of \eqref{eqpsi} represents an axisymmetric vortex centered at $(x_0, y_0)$, 
and the
second term represents a standing wave with zonal wavenumber $2$ around $(0,0)$. The amplitudes $\varepsilon_0$ and $\varepsilon_2$ are given by,

\begin{align}
 &\varepsilon_0(t)= \eta_0(1+\sin(\mu_0 t+\pi)), \\
 &\varepsilon_2(t)= \eta_2(1+\sin(\mu_2 t))
\end{align}

\noindent
Note that the intensity of the vortex and amplitude of the wavenumber 2 perturbation can vary in time such that one increases when the other decrease and vice 
versa in a very primitive representation of wave-mean flow interactions.

The specific structural forms of $F_0$ and $\Psi_2$ in our KM are given by,
\begin{align}
&F_0(r')= e^{-r'}(a\, r'+a-r'( r'+2)-2),\label{eq:psi0st} \\
&\Psi_2(r, \lambda) = r^2 e^{-r^2}\sin(2 \lambda+ { {\pi}/{2}}). 
\label{eq:psi2st}
\end{align}

\begin{figure*}
\begin{tabular}{l|l}
 Kinematic Model   &  Reanalysis Data\\ 
 \hline
 (a)  & (b) \\ 
 \includegraphics[scale=0.6]{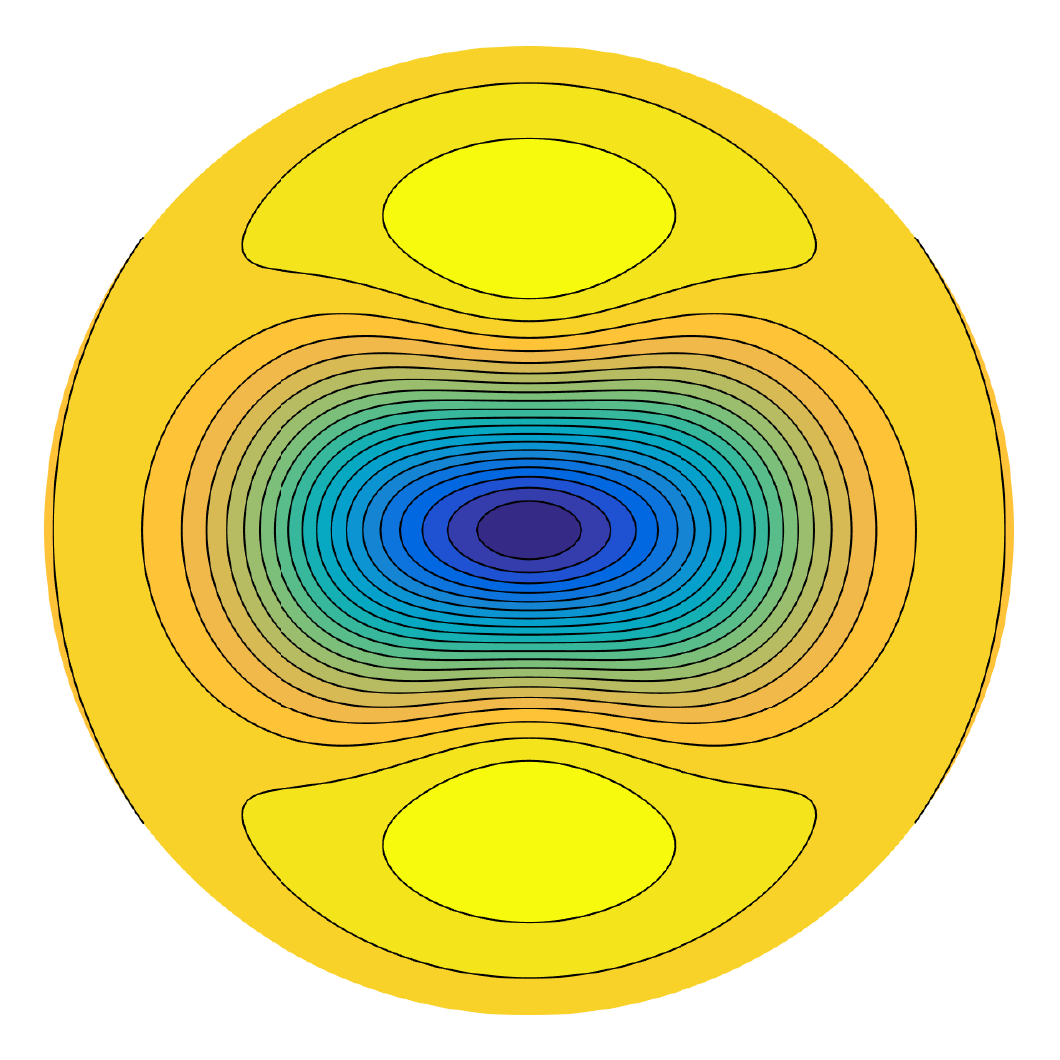} & 
 \includegraphics[scale=0.6]{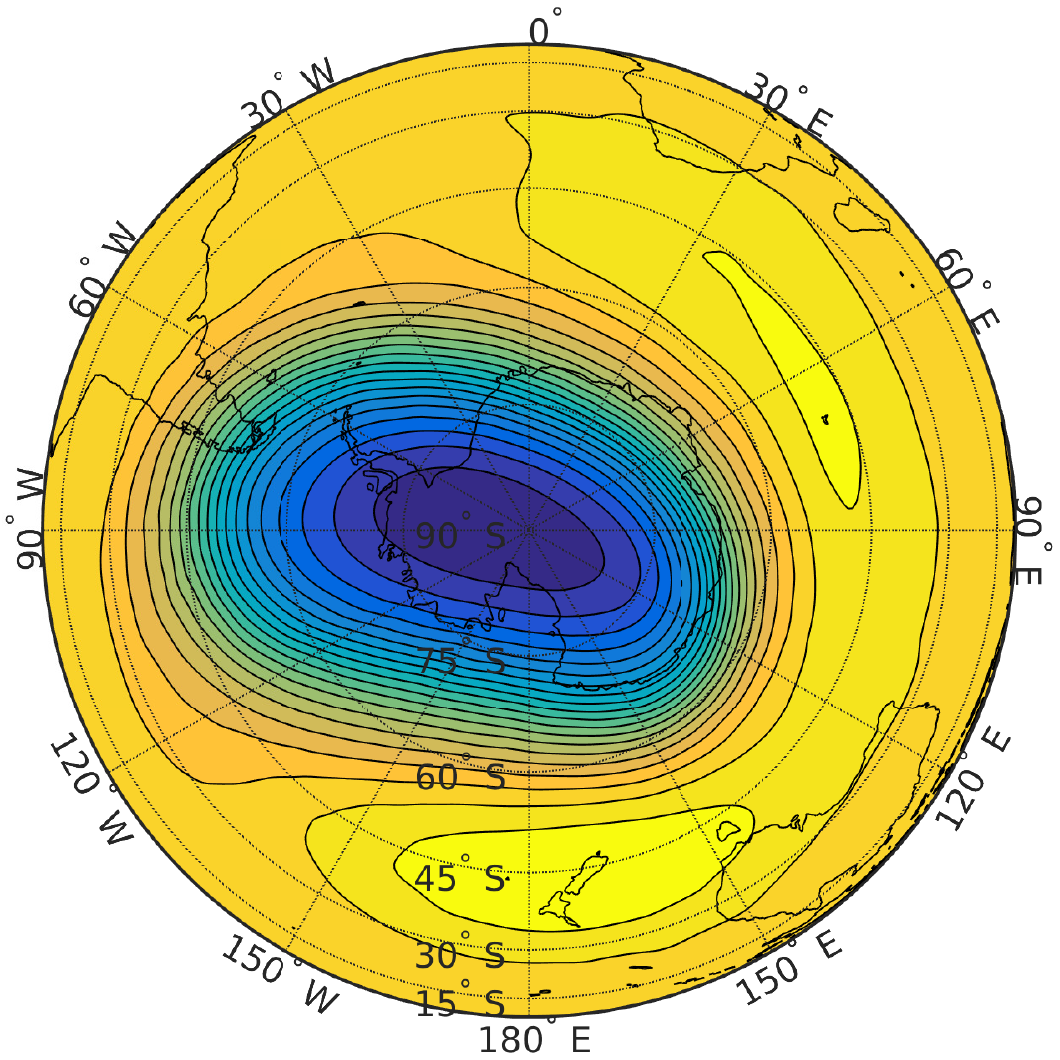}\\
 (c) & (d)\\
 \includegraphics[scale=0.3]{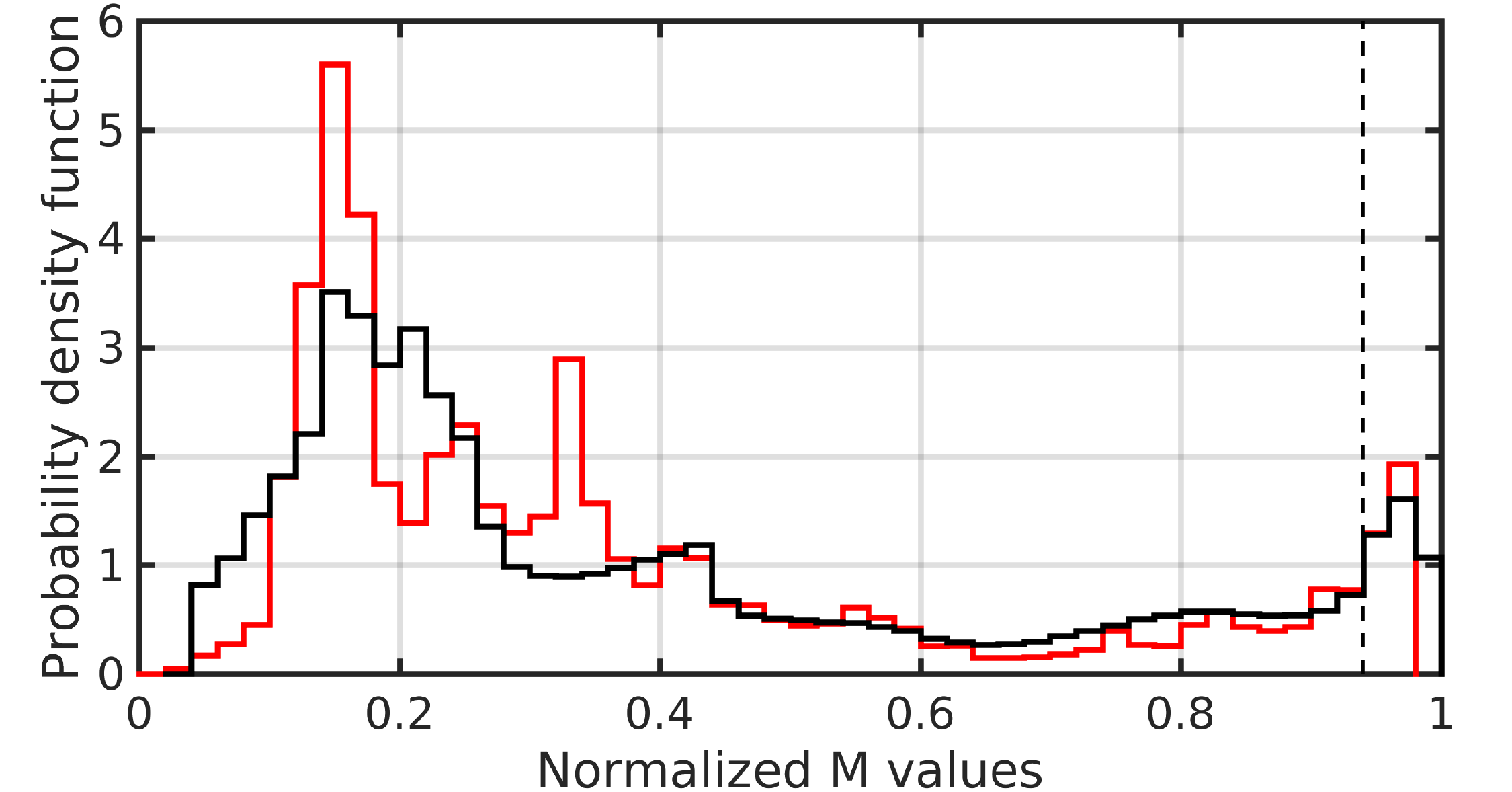}& 
 \includegraphics[scale=0.3]{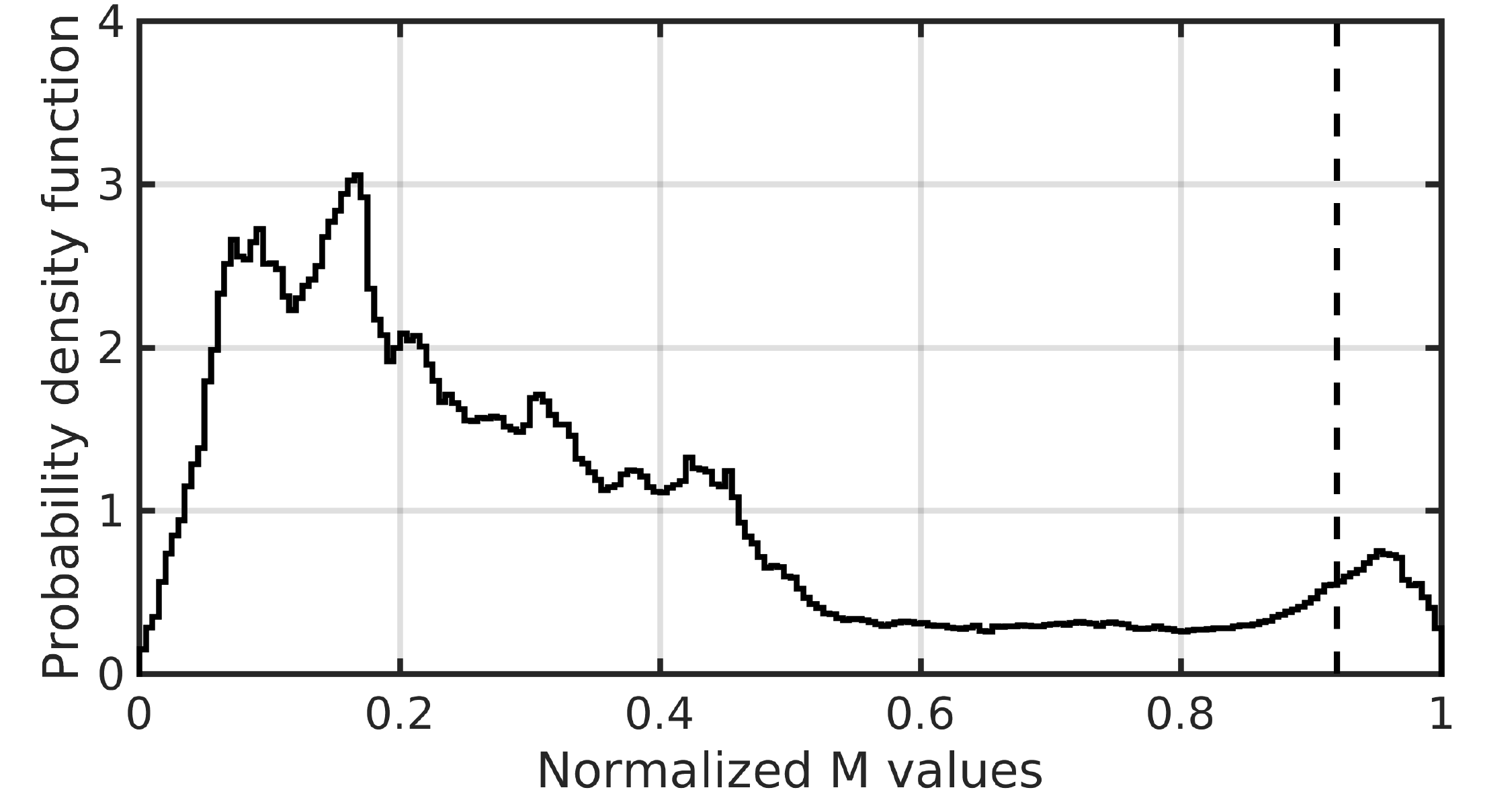}\\
 \end{tabular}
 \caption{(a) {$- \Psi$ given by equation \eqref{eqpsi}} with parameters $a=2, \mu_0=0,$ $\mu_2=2\pi/10,$ ${\eta_0=2.5, \eta_2= -1,}$ and 
axisymmetric vortex centered at the origen at $t=0$. {There are 20 contour lines between 0 (blue) and 1.4214 (yellow) uniformly distributed.} (b) 
Orthographic projection of the geopotential height field for the 10 hPa pressure level on 9 September 2002 at 00:00:00 UTC. {There are 20 contour lines 
between    27.5km (blue) and 31.3km (yellow) uniformly distributed.} (c) Probability density 
function estimate for the normalized function $M$ computed with $\tau=5$ (black line) and $\tau=15$ (red line) over a period of ten days. The area of each bar 
is the relative number of observations in the bin; hence the area under the curve is 1. The dashed line corresponds to normalized $M$ equal to 0.94. (d) 
Probability density function estimate of the normalized function $M$ ($\tau = 5$) at 850K for a ten-day period centered on 9 September 2002 . The 
dashed vertical line corresponds to 0.92.}\label{fig:pdf_streamfunction}
\end{figure*}

 \noindent
 These specific forms are inspired by the corresponding structures in the observational data for September 2002.  
Figure \ref{fig:pdf_streamfunction}(a) shows that the streamfunction in the KM for  $a=2, \mu_0=0, \mu_2=2\pi/10, {\eta_0=2.5, \eta_2= -1}$ and $t=0$ 
closely 
resembles the observed geopotential field at 10 hPa on 9 September 2002 (Figure \ref{fig:pdf_streamfunction}(b)).

The KM with the appropriate parameters can also capture the spectral distribution of observed velocities. \cite{nz} associate the jet with values of $M$ that 
are above a certain threshold found by inspection of the probability distribution function (PDF) of $M$. The PDF resembles a Gaussian distribution with a fat 
tail. We plot in Figure \ref{fig:pdf_streamfunction}(c) the PDF of $M$ for the same parameters used obtain Figure \ref{fig:pdf_streamfunction}a) for two 
different values of $\tau$.  Figure \ref{fig:pdf_streamfunction}(d) show PDFs of $M$ ($\tau = 5$) at 850K for a ten-day period centered on 9 September 2002. 
Again, the similarity between KM fields and observations is apparent.

\begin{figure*}
\begin{tabular}{ll}
 (a)&(b)\\
 \includegraphics[scale=0.7]{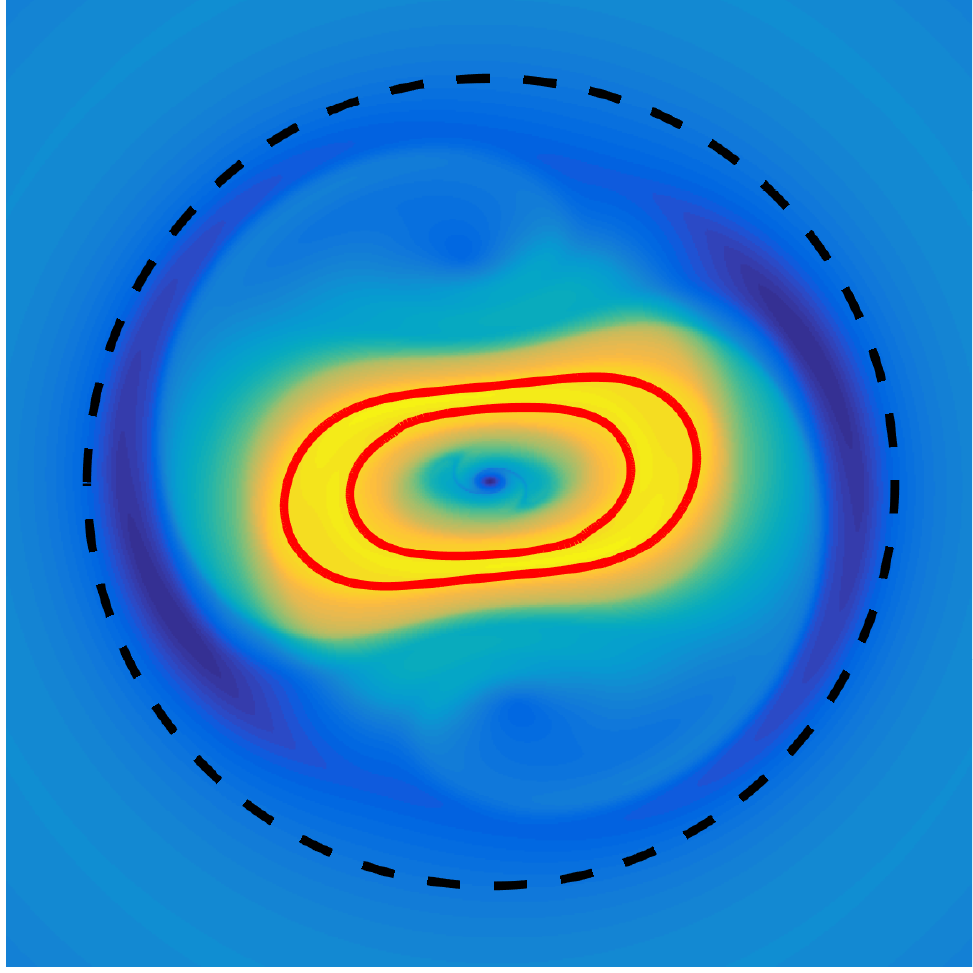} &\includegraphics[scale=0.7]{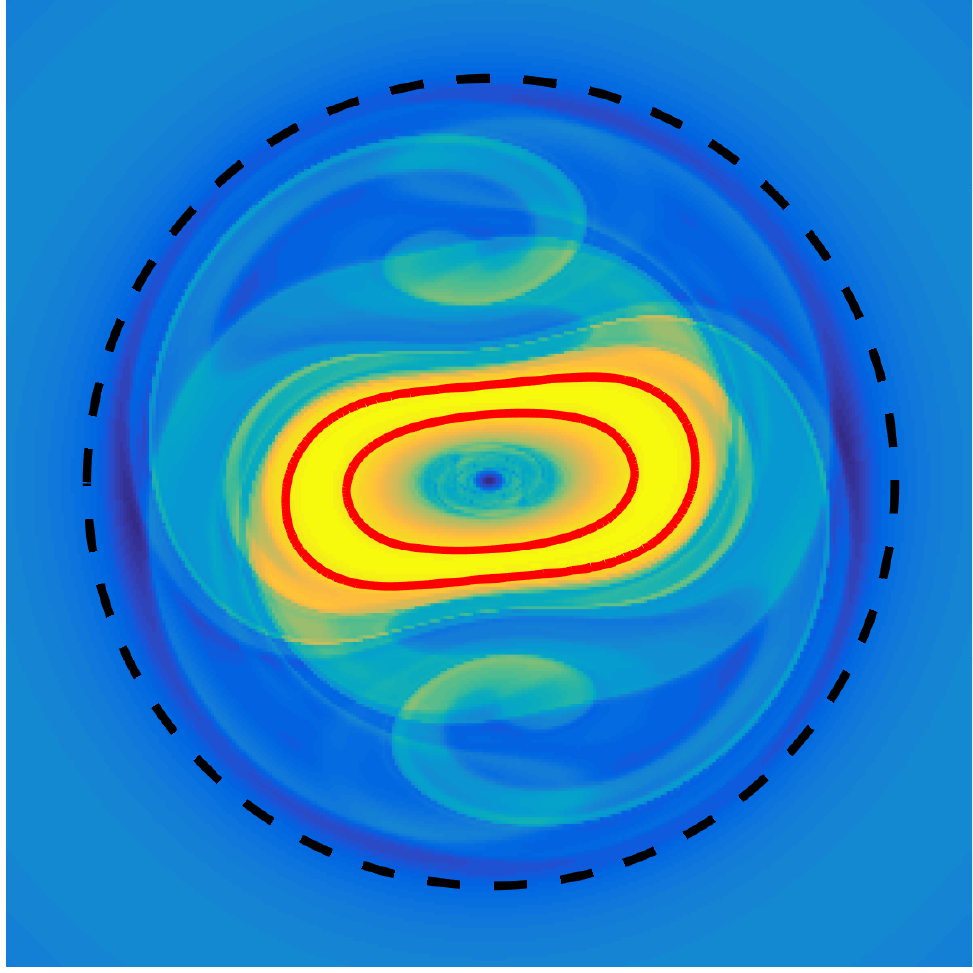}
 \end{tabular}
 \centering
\includegraphics[scale=0.7]{colormap_M_padula_big_2.pdf}
 \caption{The  {normalized} function $M$ obtained at $t = 0$ computed with $\tau = 5$ (a) and $\tau=15$ (b) from the KM 
\eqref{eqpsi} with parameters $a=2$, $\mu_0=0,$ $\mu_2=2\pi/10,$ {$\eta_0=2.5$, $\eta_2= -1$} and $(x_0,y_0)=(0,0)$. The red lines are
the contours of normalized $M$ equal to 0.94. We note the increasing richness in detail with increasing $\tau$. }
\label{fig:KMvortex}
\end{figure*}

\section{Introduction of a definition for the vortex boundary based on M}
\label{s:lagrangianstructures}

In terms of the PV field, the SPV edge on an isentropic surface has been indicated simply by an isoline determined by inspection 
\cite{manney1994a,waugh_randel99, manney2016}, {using the total ozone field \cite{Bowman93}}, or by the latitude across which the meridional gradient PV 
field is maximum for a particular longitude (\cite{mcintyre84,juckes87,alvaro2}). A vortex border region has also been defined as a domain bounded by PV 
isolines (e.g., \cite{Fairlie1988, Trounday95, ollers2002}).

The computation of PV maxima at particular longitudes can produce ambiguous results, especially when the vortex splits or is no longer around the pole as 
meridians may intersect more than one vortex edge. In order to bypass this difficulty, \cite{Nash96}  choose the location of the polar vortex boundary using a 
methodology in which PV is expressed as a function of a single variable: the equivalent latitude. For a value of PV, its equivalent latitude (ePV) is defined 
as that of the border of a polar cap with the same area as the one enclosed by the PV isoline on the isentropic surface.  The one-dimensional function ePV may 
have several local maxima.  The special PV isoline that defines the vortex edge is defined as the one where the derivative of ePV is maximum and is near the 
location of the maximum wind speed calculated along the PV isolines. \cite{nz} mentions that the equivalent latitude of the vortex edge can be defined as that 
where the ``impermeability function'' is maximum, in which this function is defined as the product of the derivative of ePV times the velocity also expressed 
as 
a function of equivalent latitude. Definitions that use the concept of equivalent latitude generally result in closed isolines of PV around local maxima of 
this field, which can be more than one as it occurs when the vortex splits.

There are also definitions that employ Lagrangian concepts. \cite{Dahlberg94} define the vortex edge as the PV contour with the minimum flux of parcels across 
it. \cite{Santitissadeekorn2010} estimates the structure of the polar vortex with a technique that finds regions with minimal external transport based on 
transfer operators. \cite{Serra2017} define the vortex edge on isentropic surfaces on the basis of Lagrangian Coherent Structures (LCSs). They show that a 
curve 
defined as the outermost elliptic LCS around the polar vortex region forms an optimal, non-filamenting transport barrier dividing the vortex core from the surf 
zone. 

We are interested in revisiting the definition of the SPV boundary in order to help with the selection of parcel trajectories that will be used in Part II to 
visualize the 3D evolution of the event. In our scenario, the boundary region of a vortex would be such that,  {(i)} it divides the core from its surroundings, 
and  {(ii)} it is free of hyperbolic trajectories and hence does not produce filaments during a certain time interval.  A region with these properties, if it 
exists, would emulate a transport barrier between the vortex core and the ``surf zone'' during a time interval to be specified. 

Let us examine the problem of finding a region with such properties in the context of the KM given by a circumpolar jet plus a standing wave 2 both with a time 
varying amplitude and the following set of parameters: $ a=2, \mu_0=0, \mu_2=2\pi/10, {\eta_0=2.5, \eta_2= -1}. $  For a given sufficiently large $\tau$, 
the 
region with values of $M$ above a sufficiently high threshold would satisfy requirement  {(i)} because trajectories will tend not to cross locations 
characterized by parcels that travel at the highest speeds throughout the interval $(t-\tau, t+\tau)$,  such as those in the jet itself.  Typically jets are 
characterised by the trajectory geometry and also in terms of particle  speed since those  in the jet travel faster than those  outside the jet.  A reasonable 
choice for the threshold is the lower limit of the fat tail in the PDF of $M$, as marked in Figure \ref{fig:pdf_streamfunction}(c) and which corresponds to the 
upper 94 per cent of the values (normalized $M$=0.94). The two isolines corresponding to this threshold are shown in red  in Figure \ref{fig:KMvortex}.  (Note 
that despite the increased richness in $M$ details with a higher $\tau$, the threshold value is practically the same.) To address requirement  {(ii)} we recall 
that \cite{mm2010,mm2012} conjectured that for sufficiently large values of $\tau$, cross-hatched patterns in a color-coded plot of $M$ can be associated with 
the intersection of unstable and stable manifolds of associated hyperbolic trajectories. It is reasonable to assume that no such structures form 
in the narrow confines of the two red curves in Figure \ref{fig:KMvortex}, at least for the values of $\tau$ used in the KM integration (5 and 15).  Hence, we 
posit that the region between the red curves represents a barrier without (or with minimum) filamentation in the interval $(-\tau, \tau)$.

\begin{figure}
\centering
 \includegraphics[scale=0.6]{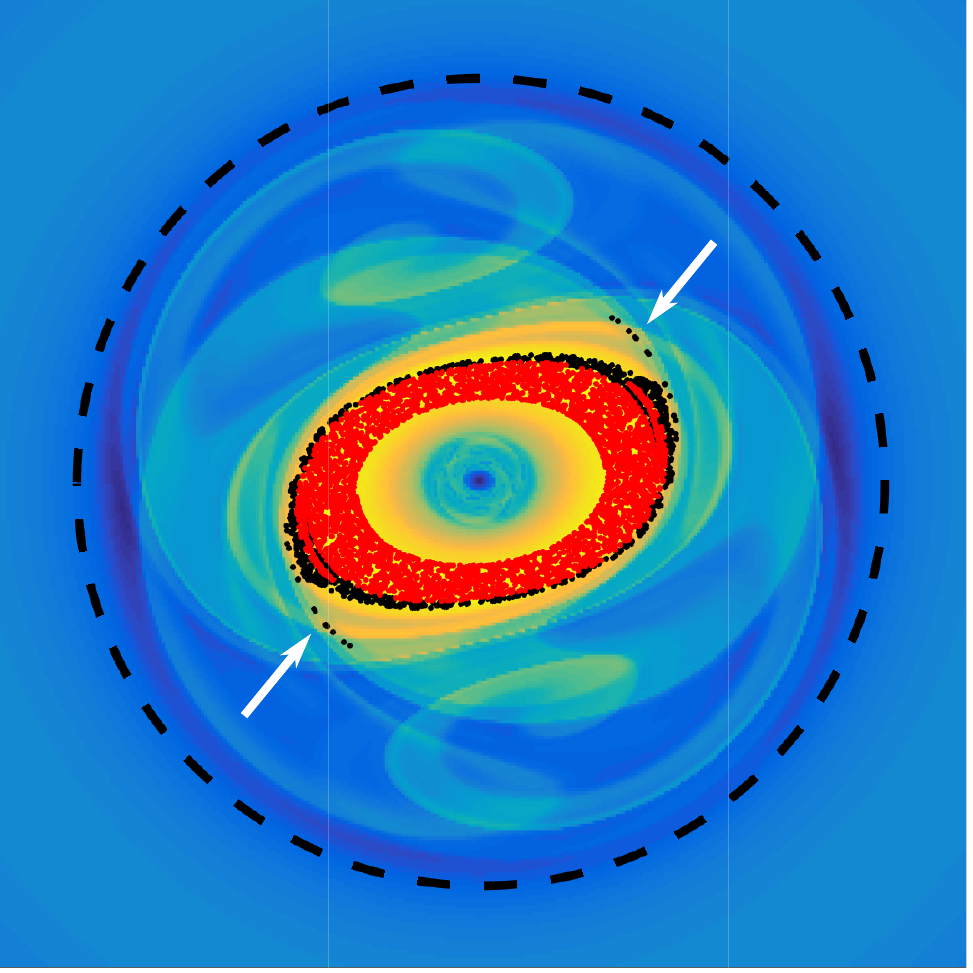}\\
 \centering
\includegraphics[scale=0.4]{colormap_M_padula_big_2.pdf}
 \caption{The function $M$ ($\tau = 15$) obtained at $t = 29$ from the KM \eqref{eq:psi0st}-\eqref{eq:psi2st} with parameters $a=2$, $\mu_0=0,$ 
$\mu_2=2\pi/10,$ {$\eta_0=2.5$, $\eta_2= -1$} and $(x_0,y_0)=(0,0)$. The red points are the position of the particles that at time $t=0$ are between the 
two red 
lines of Figure \ref{fig:KMvortex}, i.e. they have values of  {normalized} $M$ larger than  {0.94}. The black points are the position of 
the particles that have at time $t=0$ values of  {normalized} $M$ between  {0.89 and 0.94}. The white arrows highlight the hyperbolic points.} 
\label{fig:particles2DKM}
\end{figure}

To verify the plausibility of identifying  the region delimited by the isolines determined using the PDF of $M$ with a kinematic vortex boundary, we examine 
the 
parcel trajectories.  First, we select at $t=0$ parcels within the two red curves in Figure \ref{fig:KMvortex} (i.e.,  normalized $M$ larger than  
0.94), and label them with red dots.  Second, we take parcels just along and outside of the outer red curve in Figure \ref{fig:KMvortex} (i.e.,  $0.89 \leq 
\text{normalized $M$} \leq 0.94 $), and label them with black dots.  The position of the selected parcels at $t=29$ are indicated in Figure 
\ref{fig:particles2DKM}. We can see that red dots have remained within the region we are referring to as the kinematic vortex boundary.  Although most black 
dots 
have remained also close to the outer red curve,  however, we can also see two filaments of black dots associated with the hyperbolic trajectories highlighted 
by the arrows. We conclude that our procedure can provide the approximate  location of the poleward boundary of the ``surf zone''  as well as the equatorward 
boundary of the vortex interior.

A more formal approach to describing the boundary of the  jet that forms the SPV is discussed by \cite{irina}. Their approach is based on the presence of 
dynamical barriers formed by fluid parcels evolving in closed trajectories, i.e. periodic trajectories, that trap regions of fluid in their interior. 
Mathematically these objects are denoted as tori.  In 2D incompresible stationary flows, these closed trajectories are related to  level sets of a 
streamfunction that  at all times  would {resemble that} depicted in Figure \ref{fig:pdf_streamfunction}a). In these cases,  the periodic trajectories are
characterized by a  single frequency and for this reason are called 1-tori. A single closed contour of the stationary  streamfunction is an invariant set (see 
the annex A for this definition) and therefore it is a barrier to transport.  If the fluid is subjected to certain classes of time-dependent perturbations, 
some 
nonchaotic trajectories survive in the perturbed system. The conditions for this to occur are given by the Kolmogorov-Arnold-Moser (KAM) theorem 
\cite{irina,wmnpg2014}. These trajectories, which are characterized by a higher number of frequencies and are called n-tori, still have the ability to trap 
fluid in their interior, and are therefore barriers to transport. Unfortunately in fluid flows subjected to time dependent perturbations, as the ones we are 
considering in the present study,  periodic solutions related to tori cannot be obtained from the contourlines of the streamfunction. The recent paper by 
\cite{ld_ijbc} provides a way to compute tori by using the function $M$. This methodology is  firmly rooted in results of the Birkhoff ergodic theorem, which 
is 
applicable in this case since the flow represented by our KM has a Hamiltonian structure and is defined on a bounded domain (\cite{kinm}). Under these 
conditions Birkhoff ergodic theorem states that the function $M$ divided by $2\tau$, i.e. the average of  $|\mathbf{v}(t,x,y)|$ along the trajectory,  
converges 
 in the limit  $\tau \to \infty$.  The  {contours} of these limit values are  invariant sets. We next discuss how a numerical implementation of this result in 
the KM provides an approach to a 2-torus, which provides a boundary region for the vortex. 

\begin{figure*}
\begin{tabular}{l|l}
(a)&(b)\\
\includegraphics[scale=0.4]{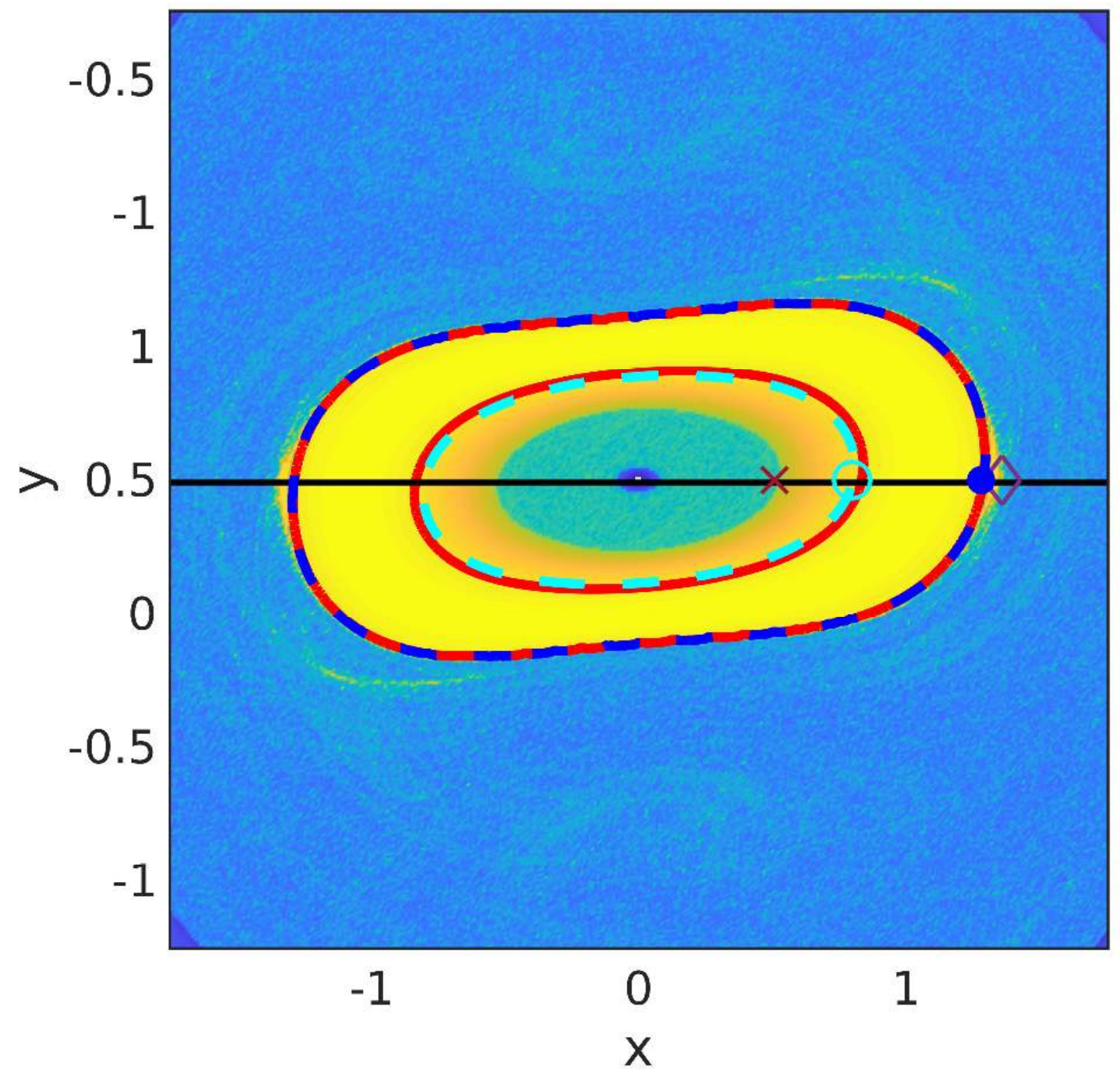}&\includegraphics[scale=0.4]{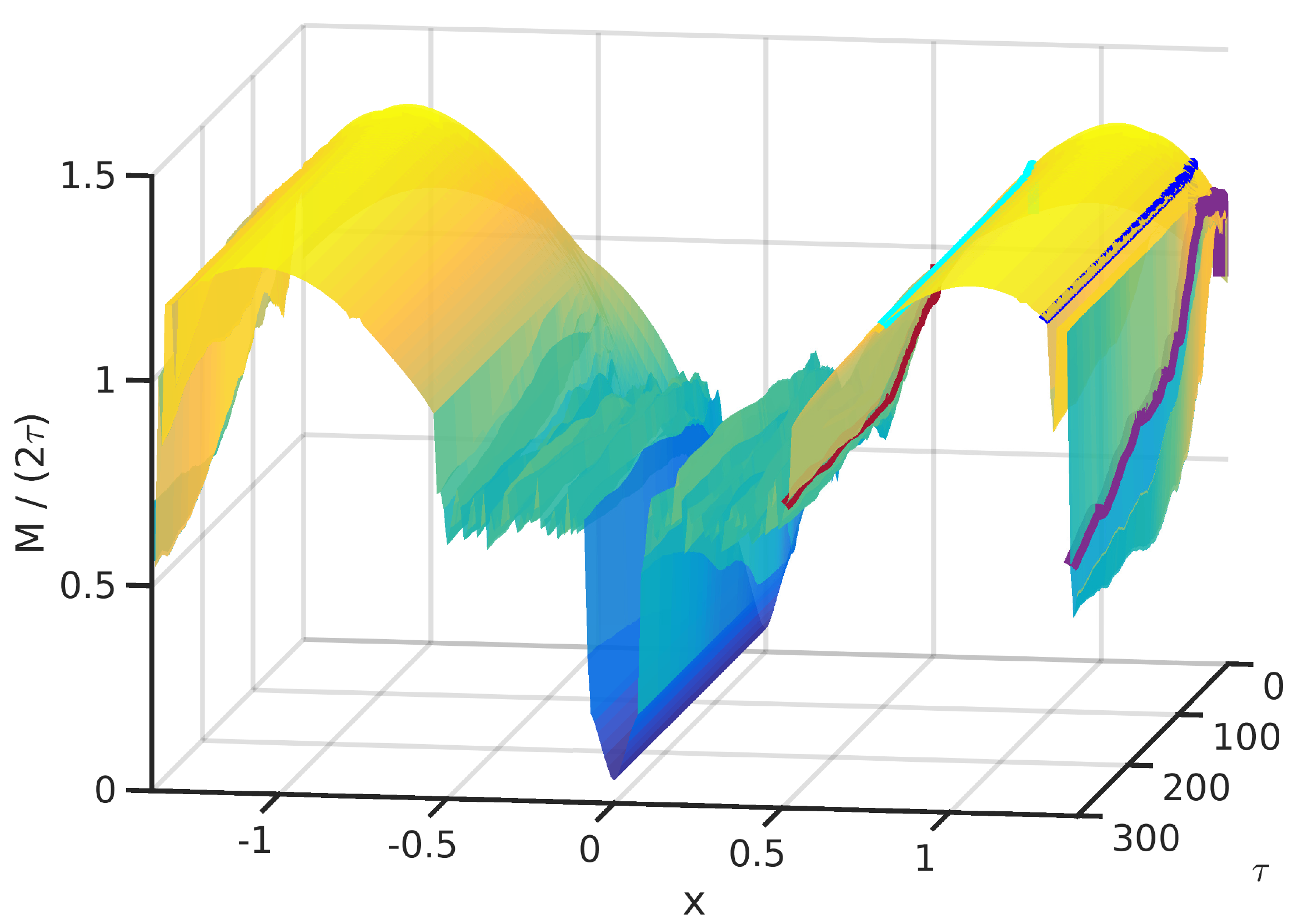}\\
 \includegraphics[scale=0.3]{colormap_M_padula_big_2.pdf} & (c)  \\
\includegraphics[scale=0.4]{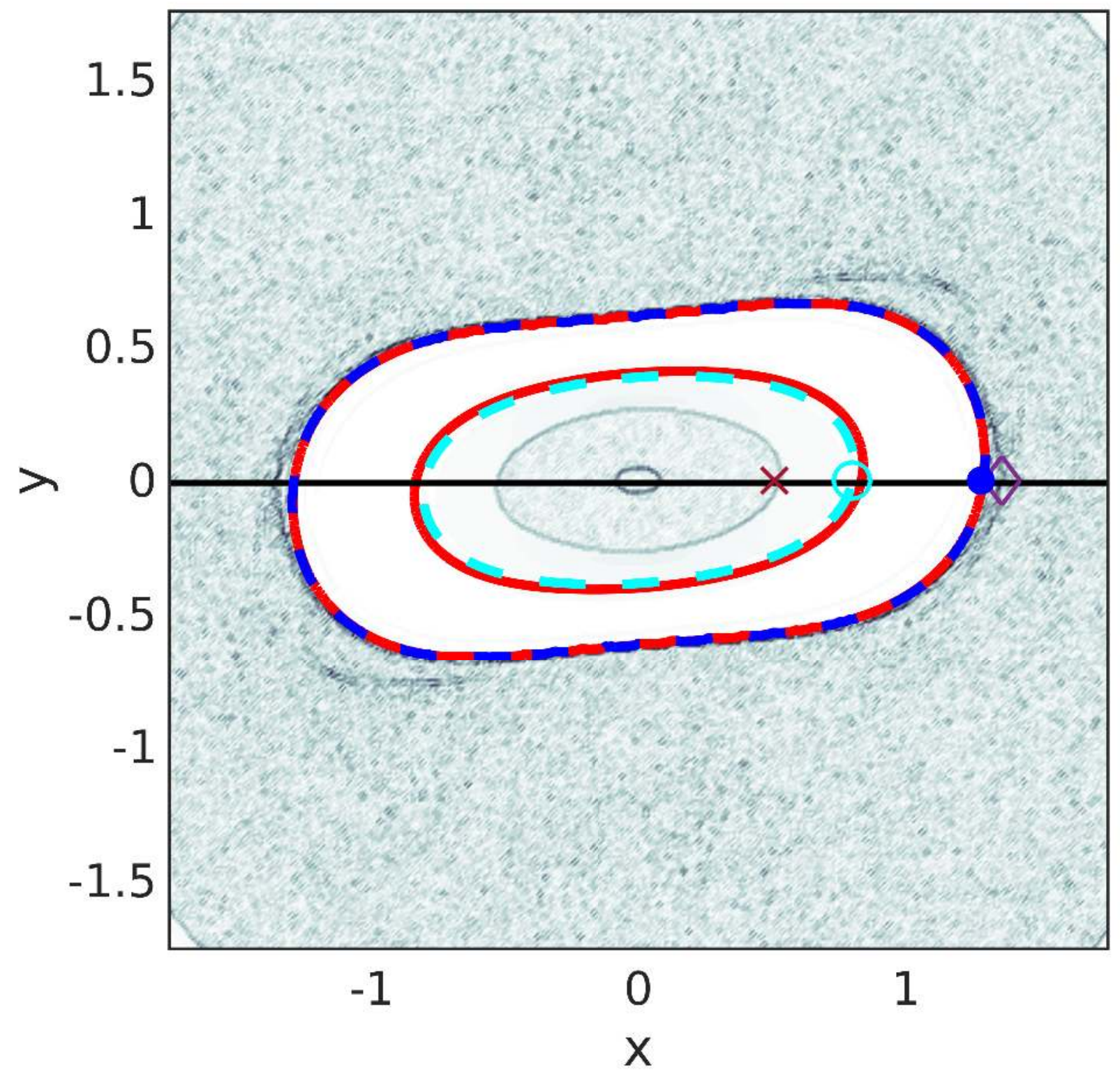}&\includegraphics[scale=0.4]{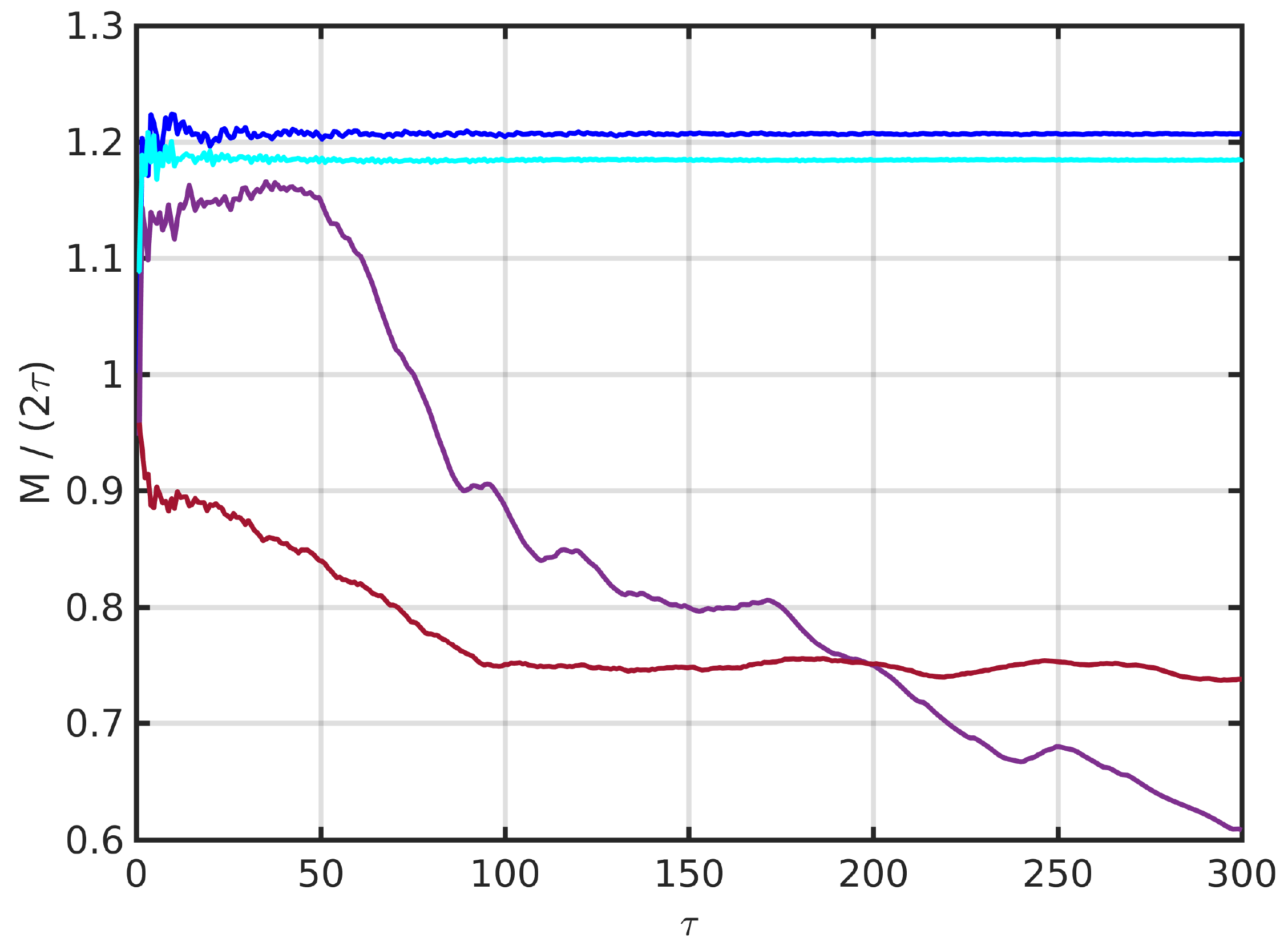}
\end{tabular}
\caption{(a) The function $M$ (top) and $\nabla M$ (bottom) obtained at $t = 0$ with $\tau = 300.$  The dashed blue and green 
contourlines mark {the approximate outer and inner 
boundaries for the 
vortex, respectively as defined by the contours of normalized $M$ equal to 0.94 (red lines). The purple diamond, blue point, cyan circle and wine cross 
in the panels of the left column correspond to points just outside, on the outer boundary, on the inner boundary,  and just inside the vortex boundary. (b) 
Values of $M$ divided by $2\tau$  in the range $\tau \in (0, 300]$ for trajectories obtained from initial conditions along the solid black line of (a). The 
colours identify the location of the initial conditions in reference to the upper panel of (a).
(c) Time evolution of $M$ divided by $2\tau$ for the trajectories from initial conditions corresponding the blue, cyan, purple and wine points in (a). Note 
that $M$ divided by $2\tau$ does not converges as $\tau$ increases for trajectories with initial conditions outside the vortex boundary.}}
\label{fig:time_evolution_M}
\end{figure*}

Let us compute the average of $M/(2\tau)$ as a function of $\tau$ for the points $(x_0,0)$ along  the 
black line in Figure \ref{fig:time_evolution_M}(a).   Figure \ref{fig:time_evolution_M}(b) shows that this average approximately converges for $\tau \sim 300$ 
if $x_0$ belongs to the interval 
{$[0.8,1.29]$}.  {In our {scenario}, we consider that these results provide a sufficient, albeit not a rigorous, evidence of convergence in the 
limit $\tau \to \infty$.} 

 {We} contrast the different behaviours obtained for points inside and outside this interval  {by showing in  different coloured lines} in Figure
\ref{fig:time_evolution_M}(c) the variation {of $M/ (2\tau)$} with $\tau$ for initial conditions {$(0,1.36)$, $(0,1.29)$, $(0,0.8)$  and $(0,0.51)$.
Starting at $(0,1.36)$ }  the purple line  in Figure \ref{fig:time_evolution_M}(c) (see also purple point in Figure \ref{fig:time_evolution_M}(a))
 reveals that convergence is not reached for $\tau\sim300$,  and hence contourlines 
do not correspond to invariant sets.  Similarly, convergence is not obtained in the interior part of the vortex starting at  {$(0,0.51)$}, according 
to the wine 
line in Figure \ref{fig:time_evolution_M}(c)  (see also the wine  point in Figure \ref{fig:time_evolution_M}(a)). {Starting at  {$(0,1.29)$} the blue 
line 
reveals that convergence is approached for 
$\tau\sim 300$. We can see in Figure \ref{fig:time_evolution_M}(a) that the blue contour marks an approximate outer edge for the vortex. Similarly, the green 
line 
obtained starting at  {$(0,0.8)$},  shows the convergence for  $\tau\sim 300$. Figure \ref{fig:time_evolution_M}(a) also shows that the {cyan} 
contour marks an 
approximate edge for the vortex interior.  
Contourlines taken on $M/(2 \tau)$, when this has converged in $\tau$, do correspond to invariant sets. Movie S1 in the supplementary material shows the  
evolution of particle trajectories placed on and outside a 
2-torus.} The coincidence between the outer and innermost invariant sets with the contours shown   in   Figure   \ref{fig:KMvortex} (b) corresponding to the 
values of $M$ given by the PDF, confirms that the heuristic approach discussed above can be justified on the basis of concepts from ergodic theory. The 
rigorous 
mathematical justification, however, can be obtained only in periodic or quasi periodic  cases \cite{MW99,susuki}.  This is because  the Birkhoff ergodic 
theorem has not been proven for aperiodically time dependent velocity fields. 

Nevertheless, there is computational evidence that this methodology is valid more generally.  The supplementary material includes a plot similar to Figure 
\ref{fig:KMvortex}(a), except for the addition of a small ($5\%$) random perturbation to the amplitude of wave 2  in the KM.  The figure shows similarities 
between the vortex edge defined by the PDF  in  the periodic and random perturbation cases. Caution must be exercised in assigning a general character to this 
result because \cite{mwcm2013} discuss a example in which for sufficiently long $\tau$, tori-like structures are broken in the case of aperiodic forcing, even 
for amplitudes one-order of magnitude smaller than those used in the periodic forcing.

\begin{figure*}
\begin{tabular}{cc}
(a) Kinematic Model & (b) Reanalysis data\\
\includegraphics[scale=0.7]{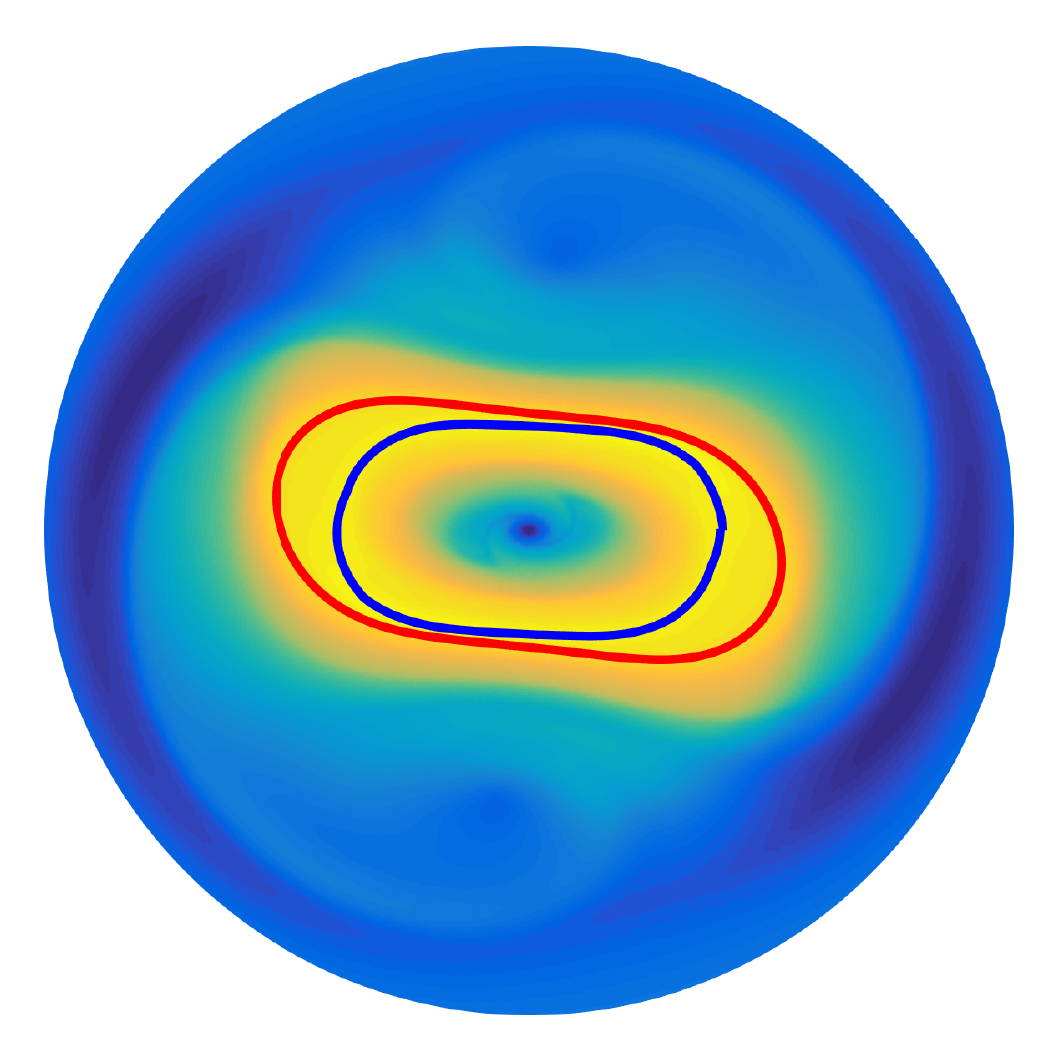} &
\includegraphics[scale=0.7]{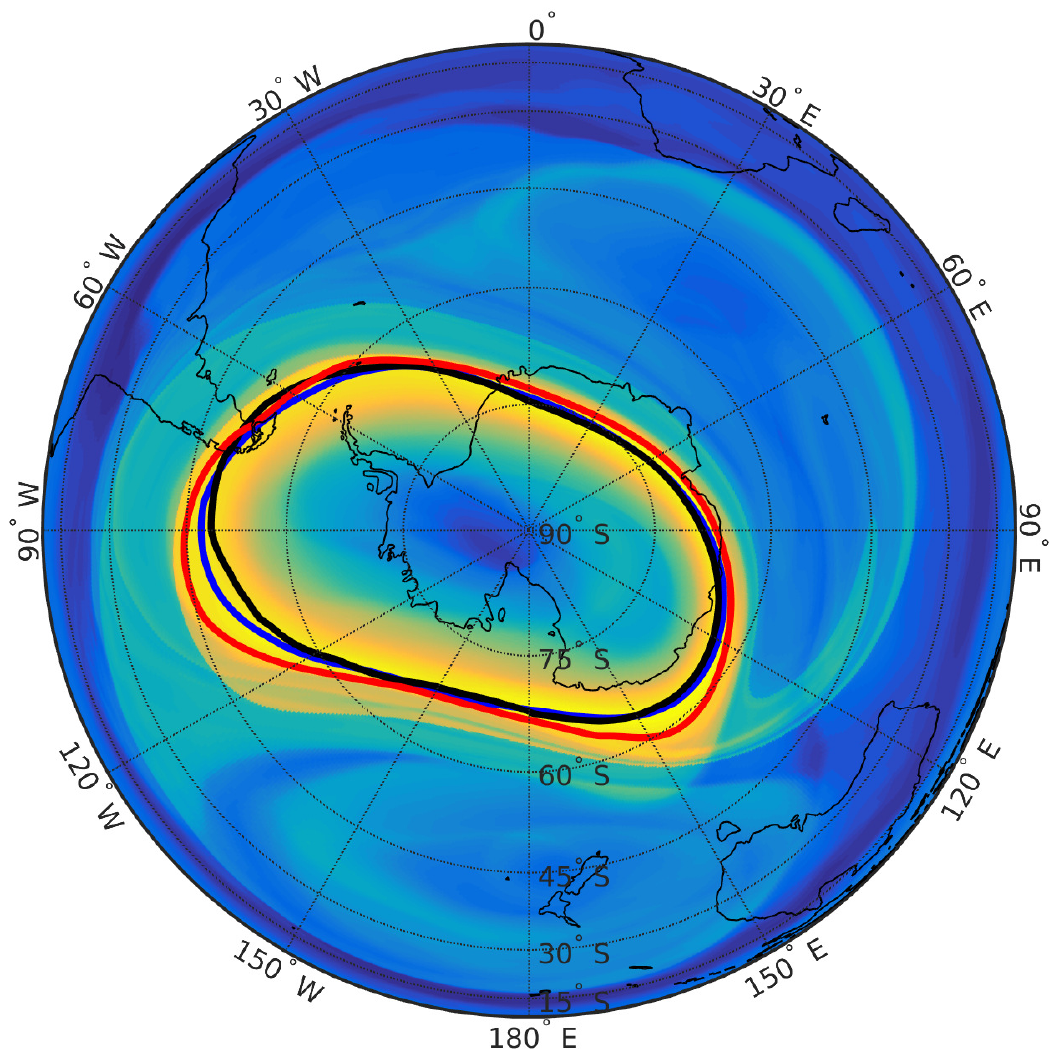}
\end{tabular}
\centering
\includegraphics[scale=0.5]{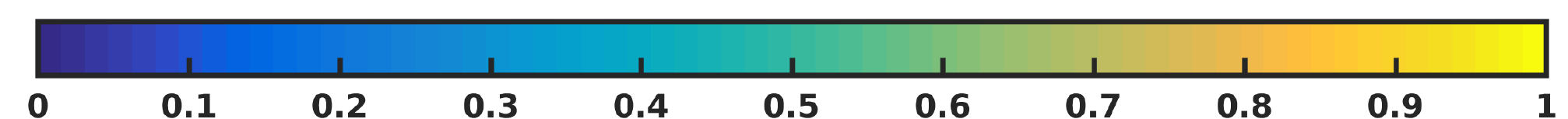}
\caption{On a background depicting the function $M$ ($\tau=5$) at {$t=5$} (a) and 9 September 2002 at 850K (b), the colored lines correspond to different 
ways to identify the vortex edge.  
Blue. Maximum of $M$ (eq. \ref{M}) for each longitude. 
Red: Outer contour of normalized $M$ equal to 0.92. 
Black (in (b) only): Maximum of the PV gradient ($\nabla PV$) for each longitude
All curves were smoothed by application of a running average of one degree of latitude (20 points).}
\label{fig:vortex_edge_data}
\end{figure*}

Thus, we have three candidates for the determination of the vortex edge.  These are listed below, to be applied on isentropic surfaces as well as on surfaces 
of 
constant height.

\begin{itemize}
\item[(i)] Maximum of $\nabla PV$ for each longitude,
\item[(ii)] Maximum of the function $M$ (eq. \ref{M}) for each longitude,
\item[(iii)]  {Outer most }contour of $M$ for  {a boundary obtained }  by inspection of the PDF (our proposal).
\end{itemize}
\noindent
The colored lines in Figure \ref{fig:vortex_edge_data} show the curves obtained with the different definitions of the vortex edge for 9 September 2002 at 850K. 
On this day and isentropic surface, the SPV was circumpolar albeit elongated and a broad filament had formed in the $130^\circ$ E - $150^\circ$ E sector. 
According to Figure \ref{fig:vortex_edge_data}(b), there are particles outside the region bounded by the lines (i)-(iii) that after 5 days remain inside the 
vortex edge.  In Part II we will be particularly interested in the behaviour of these particles, as they will tend to be ``peeled off'' the vortex along 
filaments. This can be seen in the video {S2} provided in the supplementary material.

\section{Criterion for vortex splitting}
\label{s:criterion}

In this section we examine the vortex splitting.  The function $M$ at 850K computed from reanalysis data for 24 September 2002, just at the time of the 
observed 
split before the vortex breakup, is shown in Figure \ref{fig:M}(c). We can obtain a highly similar structure in the KM by properly selecting the model 
parameters using a strong wave 2 increasing in amplitude and a substantial contribution from a vortex with a center displaced from the pole. 

\begin{figure*}
\begin{tabular}{ll}
(a)&(b) \\
\includegraphics[scale=0.4]{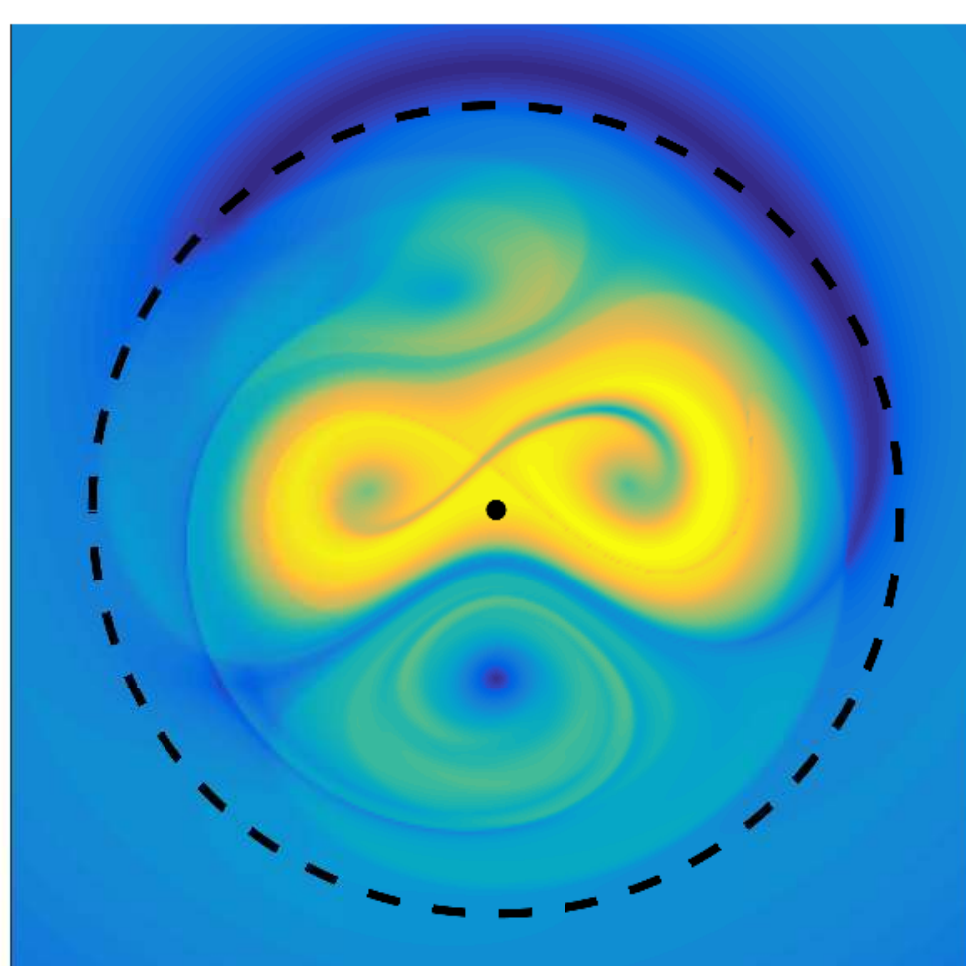}\includegraphics[scale=0.4]{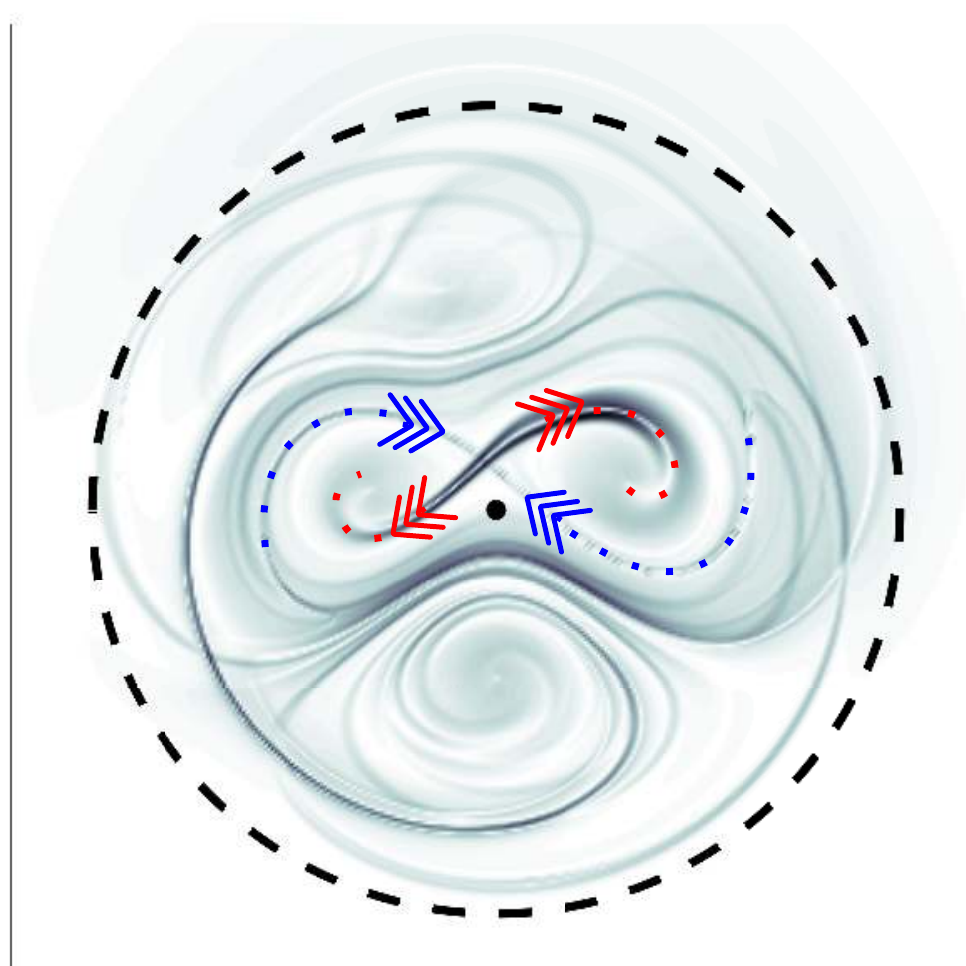}&
\includegraphics[scale=0.4]{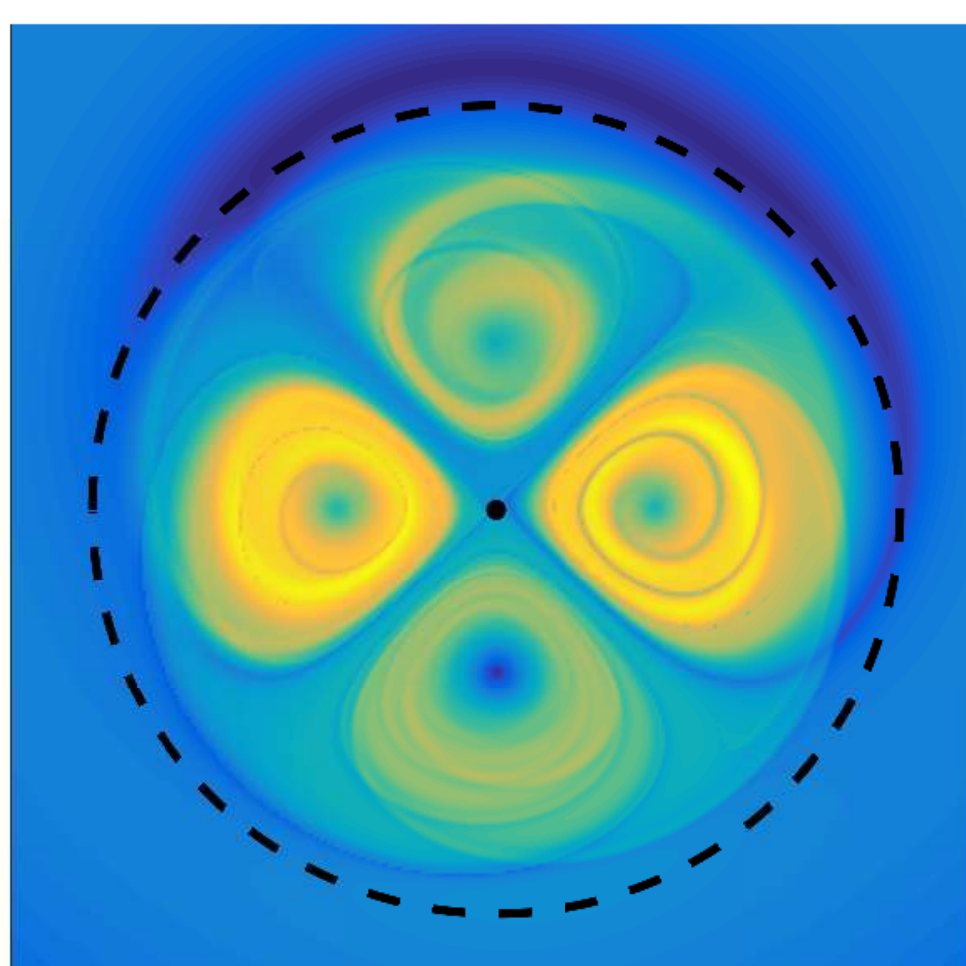}\includegraphics[scale=0.4]{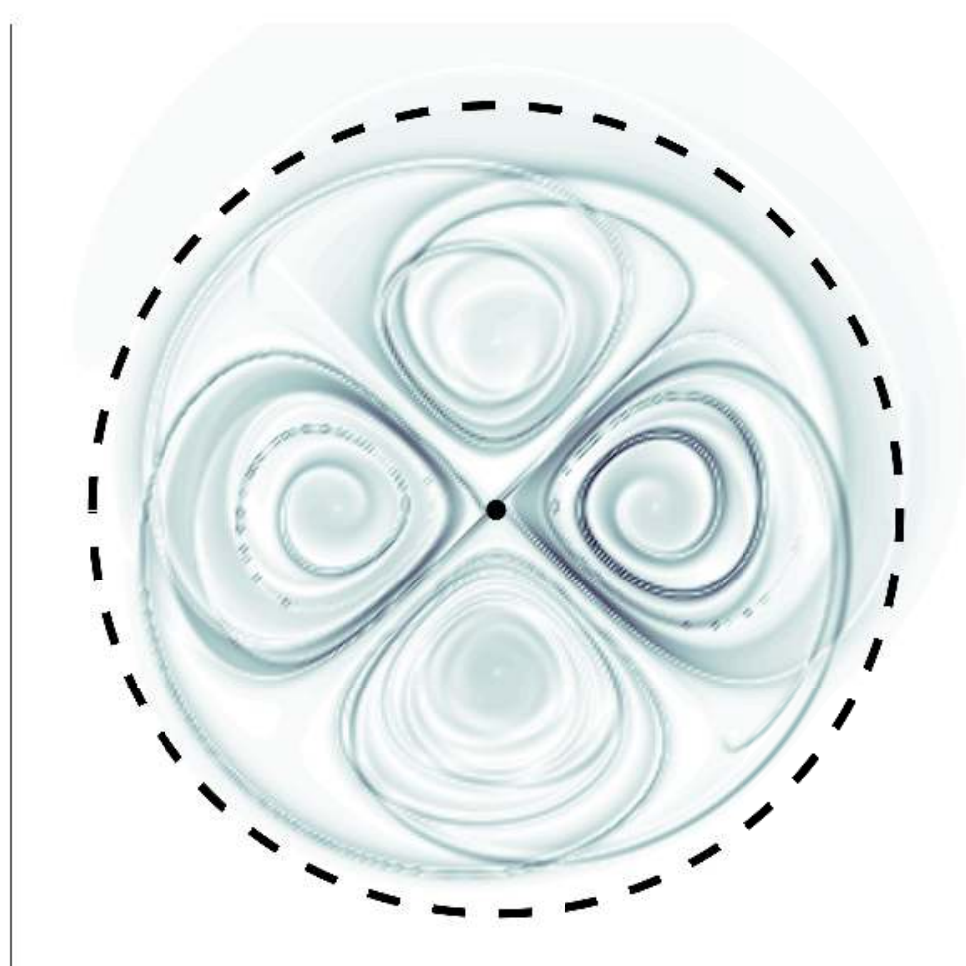}\\
(c)&(d) \\
\includegraphics[scale=0.4]{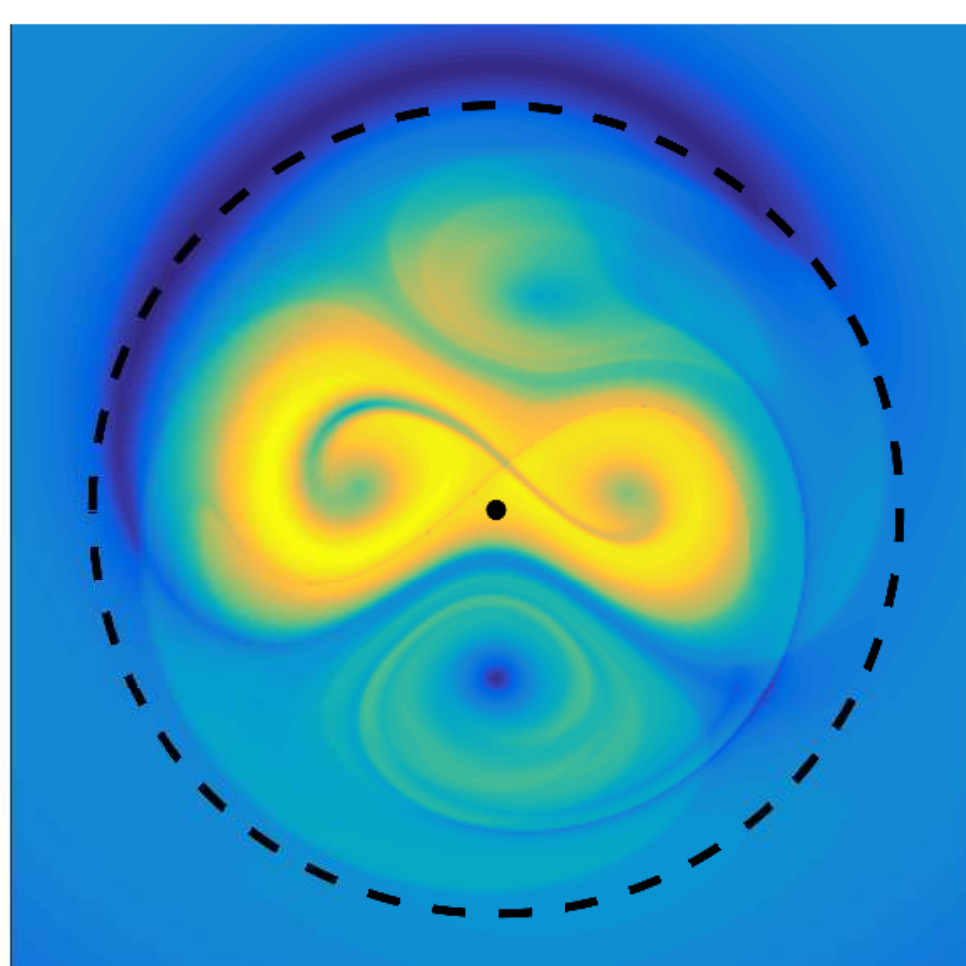} 
\includegraphics[scale=0.4]{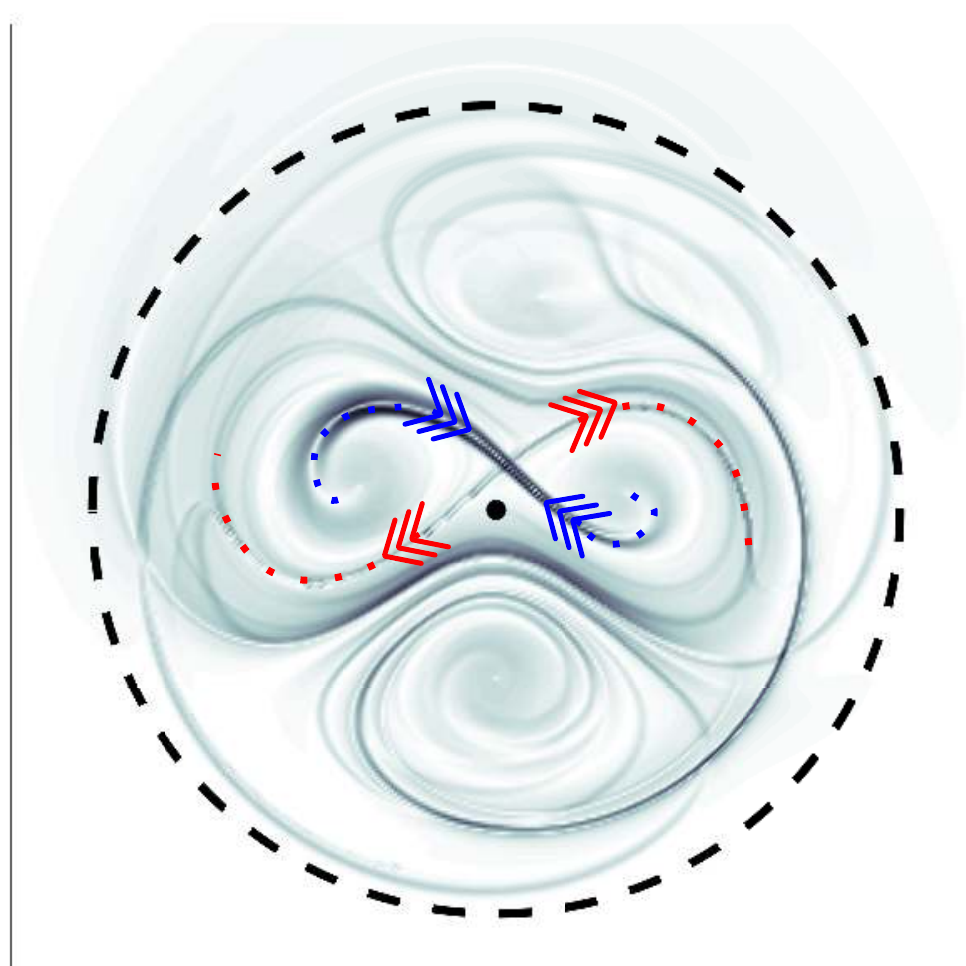}&
\includegraphics[scale=0.4]{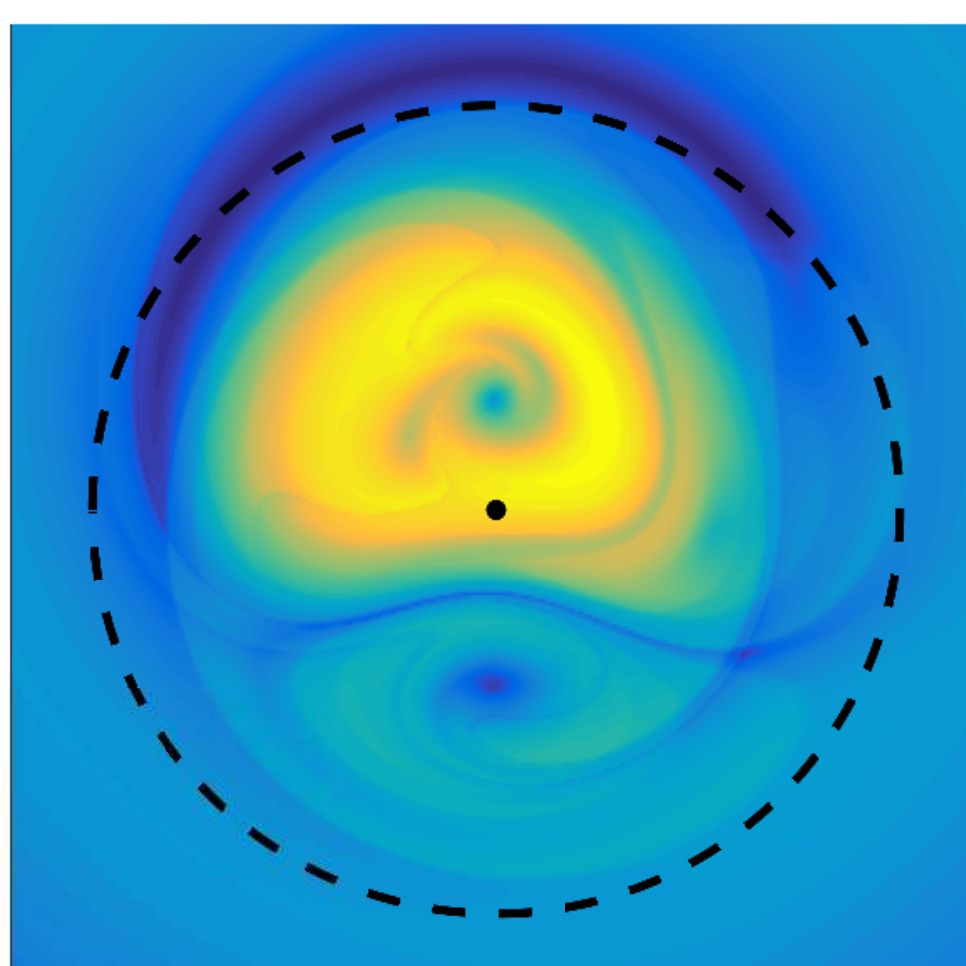} \includegraphics[scale=0.4]{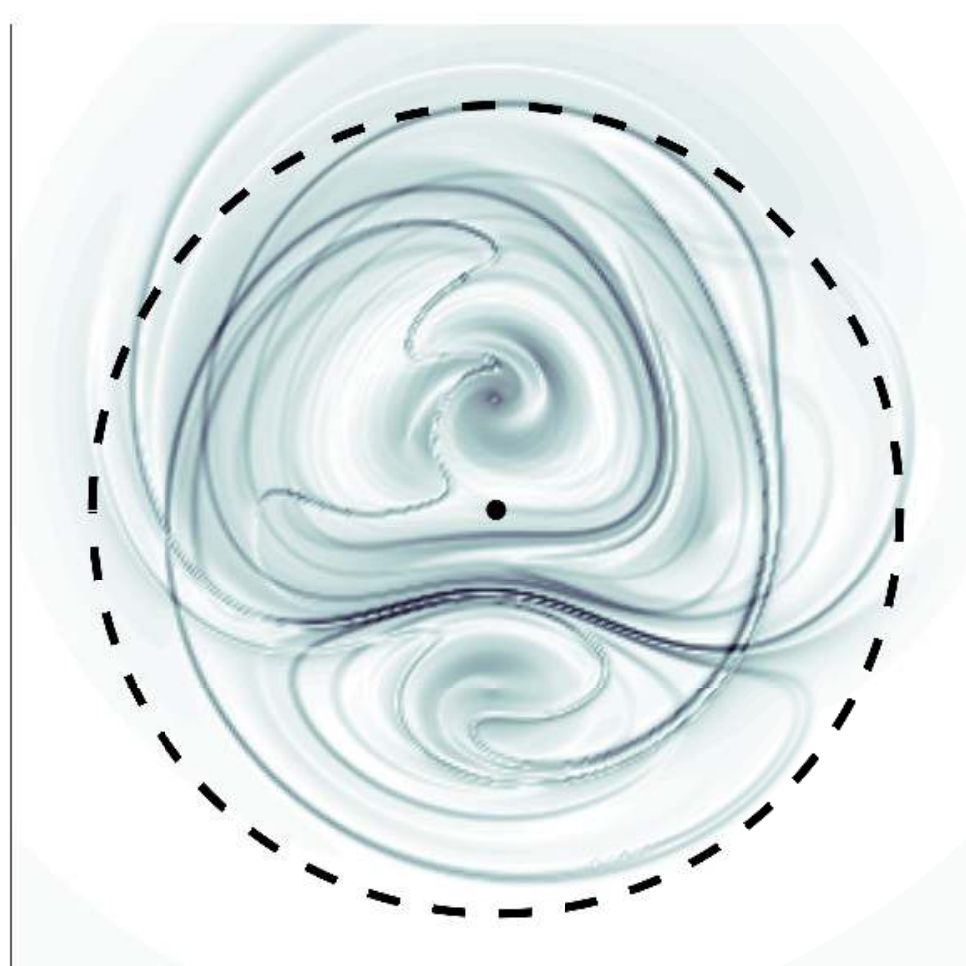}\\
\multicolumn{2}{c}{\includegraphics[scale=0.5]{colormap_M_padula.pdf}}
\end{tabular}
 \caption{The function $M$ ($\tau = 15$) and  $\nabla M$ obtained from the KM \eqref{eq:psi0st}-\eqref{eq:psi2st} at {earlier time (left two columns) and later 
time (right two columns)} with parameters $a=2$, 
$\mu_0=2\pi/50,$ $\mu_2=2\pi/30,$ {$\eta_0=2.5$,  $\eta_2= -1$} and $(x_0,y_0)=(0,0.75)$. The pole is marked with a black dot and the equator with a 
dashed black line. A pinched vortex in (a) and (c) splits into two vortices in (b), or merges into one in (d). {The 
blue and red arrows in panels (a) and (c) over  $\nabla M$ show the position of the stable and unstable manifolds, respectively in each case. This 
configuration agrees with the sketch of the structures shown in figure \ref{fig:scheme}, where the particles are organized into two vortices or in a single 
vortex. }}
\label{fig:KMpiching}
\end{figure*}

The left panel in Figure \ref{fig:KMpiching}(a) shows the function $M$ from the KM at $t = 3$ with  $a=2$, 
{$\mu_0=2\pi/50,$ $\mu_2=2\pi/30,$ $\eta_0=2.5$, $\eta_2= -1$} and $(r_0,\lambda_0)=(0.75,\pi/2)$. {At $t=3$, the KM with this set of parameters 
models a 
circular vortex displaced from the pole whose intensity is decreasing with time, and a wave 2 centered at the pole  whose amplitude is increasing with 
time. Also note that the cyclonic vortex lies above an anticyclonic lobe of    wave 2. } A hyperbolic trajectory can be 
clearly seen at a 
location close to the pole in the plot of $\nabla M$ displayed in the 
right panel of the figure. The global structure of the manifolds {depends not only} on local aspects in the neighbourhood of a hyperbolic trajectory, but 
also 
on large-scale aspects of the perturbation.  At a later time, $t=9$, this vortex splits such as shown in  Figure  \ref{fig:KMpiching}(b). In this case the 
perturbation dominates and the splitting occurs.

Not all evolutions of a pinched SPV result in splitting.  {At $t=72$, the KM with the same parameters as before models a 
circular vortex displaced from the pole whose intensity is increasing with time and a wave 2 centered at the pole  whose amplitude is decreasing with 
time. In this case, the 
pinching structure in Figure  \ref{fig:KMpiching}(c) evolves into one vortex as shown in Figure  \ref{fig:KMpiching}(d).}

\begin{figure*}
\begin{tabular}{lll}
(a) & (b) & (c) \\
\includegraphics[scale=0.45]{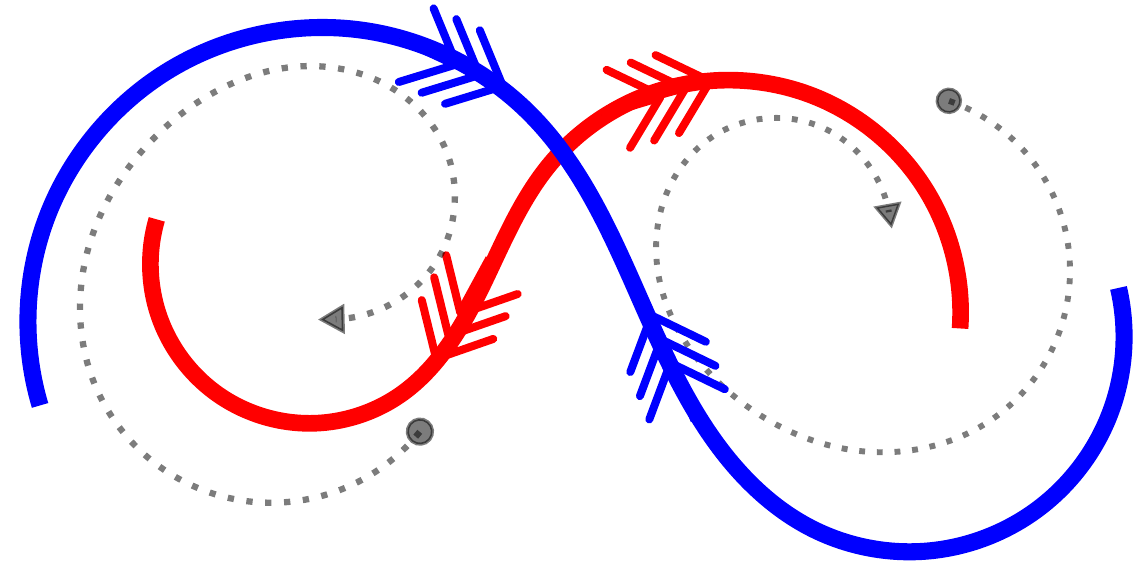} & \includegraphics[scale=0.45]{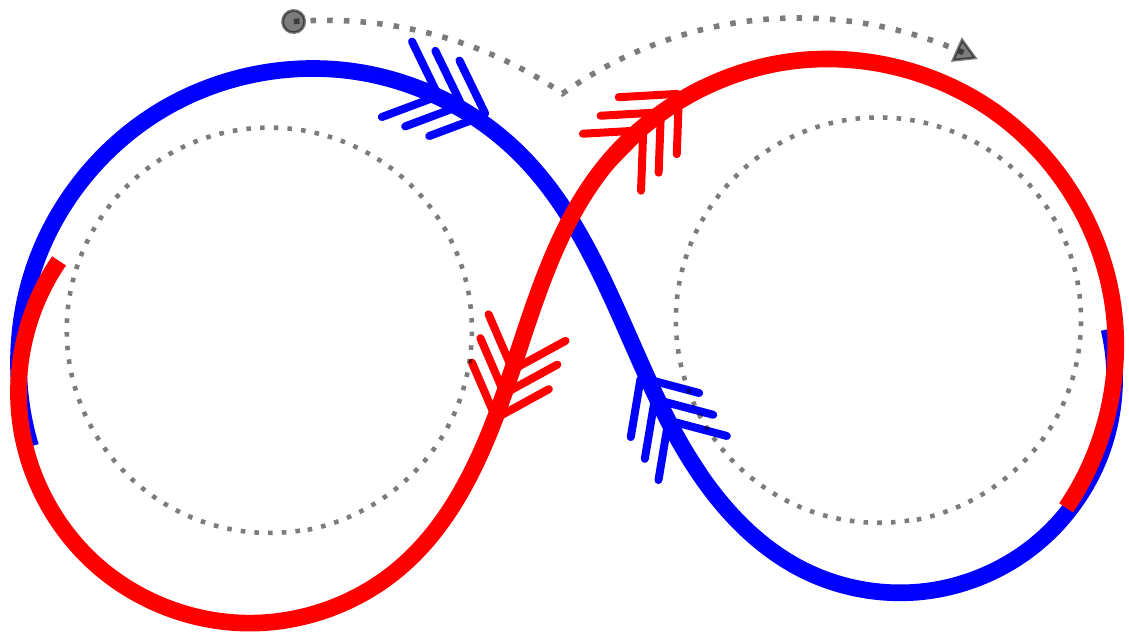} & \includegraphics[scale=0.45]{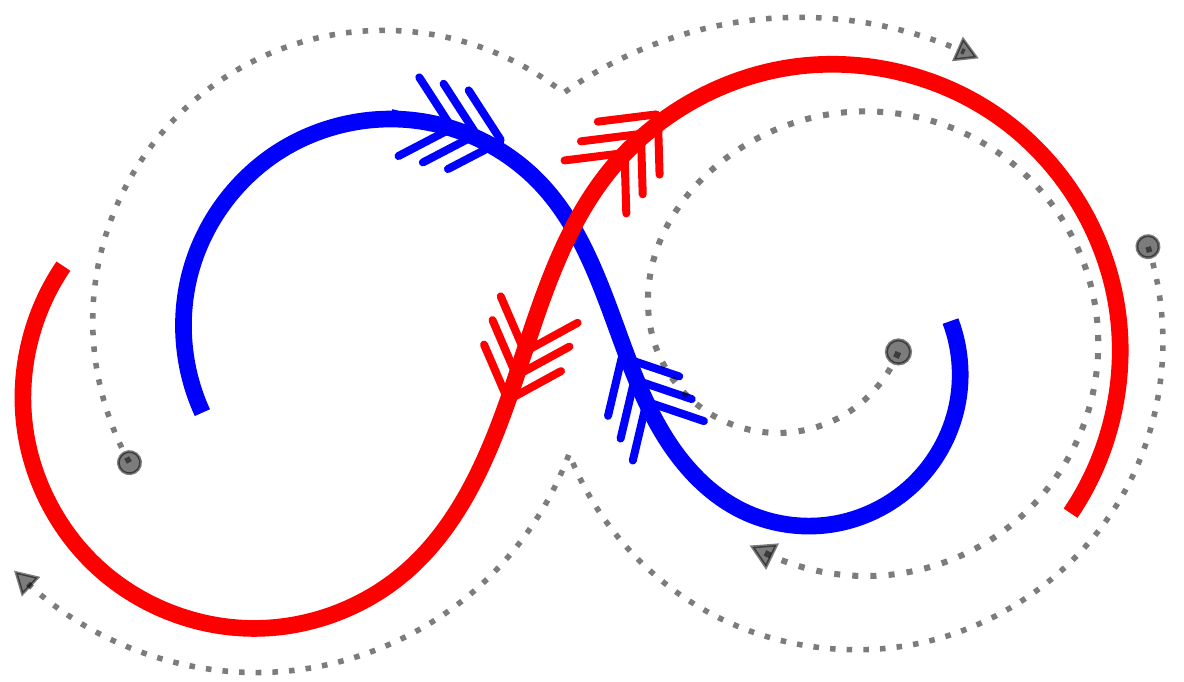} 
\end{tabular}
 \caption{Sketch showing configurations of stable (blue line) and unstable (red line) manifolds according to which one can expect at later times either vortex 
splitting (a) or no splitting (b) and (c). {The grey dashed 
lines shown schematically the trajectory of particles in the vicinity of the manifolds.}}
\label{fig:scheme}
\end{figure*}

An examination of the structure of the evolving manifolds intersecting at the hyperbolic trajectory can provide a criterion on whether the split will occur or 
not. Figure \ref{fig:scheme} is a schematic of the three possible situations for the manifold structures: (a) the unstable manifold (in red) is in the inside 
part that delimites the stable manifold (in blue) and no intersection points are detected in the {period $(t_0-\tau, t_0 +\tau)$}, (b) the  two manifolds  
are 
almost symmetric with respect to the observed intersection point opposite to the hyperbolic trajectory of reference, and (c) the stable manifold is in the 
inside part that delimits the stable manifold and no intersection points are detected in the {period $(t_0-\tau, t_0 +\tau)$}. An analysis of particle 
trajectories 
in these three instances reveals that in case (a) the structure causes the particles to organize in two vortices, while in cases (b) and (c), after pinching 
the 
particles organize in a single vortex. {In the KM with the parameters selected, configuration (a) corresponds to $t=3$, (b) corresponds to a vortex 
centered at the pole, and (c) to $t= 72$.} We will show in Part II that these different structures formed at different levels in the stratosphere in a way 
consistent with the presence or absence of the vortex splitting in September 2002.

\section{Summary and Conclusions}
\label{s:summary}

Our goal in the present two-part paper is to provide a Lagrangian perspective of the final southern warming in {spring of} 2002, during which the stratospheric 
polar vortex 
(SPV) experienced a unique splitting. In this Part I, we presented our methodology and concentrate on the fundamental processes for filamentation 
and ultimately for vortex splitting.

We have worked in the idealized context of a Kinematic Model (KM). The KM aims at emulating the behavior of the longest planetary waves on an isentropic 
surface as obtained from the reanalysis data, and allows for detailed parametric analysis. Previous studies had demonstrated that one such a model can be 
defined and applied to examine the mechanisms responsible for complex fluid parcel evolution during the polar vortex filamentation and breakdown. One of our 
objectives with the KM was to explore a kinematic definition for the SPV boundary that could be justified, at least in some cases, with arguments that go 
beyond 
heuristic considerations based on the potential vorticity field as is generally done. The second objective was to gain insight into the role of Lagrangian 
structures in the SPV splitting. 

It was argued that a kinematically meaningful definition of the SPV boundary is the region bounded by the contours of the Lagrangian descriptor known as the 
function $M$ corresponding to threshold values determined by inspection of probability distribution functions (PDFs) of $M$. In the context of the KM, this 
definition of vortex boundary was linked to rigorous results that identify a 2-torus with contours on which the values of $M/(2\tau)$ converge in $\tau$. The 
outer boundary of this region provides the approximate location of the poleward boundary of the surf zone. We provided examples of how this boundary relates to 
the formation of filaments in the surf zone. We will use this definition in Part II for the selection of particles whose trajectories will illuminate the SPV 
behavior in September 2002 at several levels. 

Finally, we used the KM to explore different behaviours of the SPV.  In particular, we examined two cases in which a pinched vortex evolved into two very 
different structures.  In one of them the vortex split completely, while in the other case the vortex returned to a less disturbed configuration.  An 
examination of the function $M$ for these cases suggested a criterion to anticipate the vortex behavior.  We shall apply this criterion in Part II to justify 
why the SPV split at some levels in September 2002, while it remained as one in other levels.

\appendix
\section{Annex}\label{A:concepts}

There exist different dynamical objects that support the qualitative description of particle time evolution. 
This qualitative description is based  on Poincar\'e's ideas, and consist of determining geometrical structures  
that define regions where particle trajectories have qualitatively different behaviours. The boundaries between these regions are dynamical barriers. These 
geometrical structures provide a template for a specific velocity field and emphasize the essential transport features associated with it.  

{\em Hyperbolic trajectories} are one type of recognisable dynamical features in flows. These are special trajectories, in the neighbourhood of which air 
masses 
are elongated along the unstable direction and compressed along the stable direction.  An important feature here is that two parcels placed at nearby locations 
close to a hyperbolic trajectory may evolve in time quite differently, separating  at exponential rates. 

\begin{figure}
\centering
\begin{tabular}{ll}
(a) & (b)\\
\includegraphics[scale=0.45]{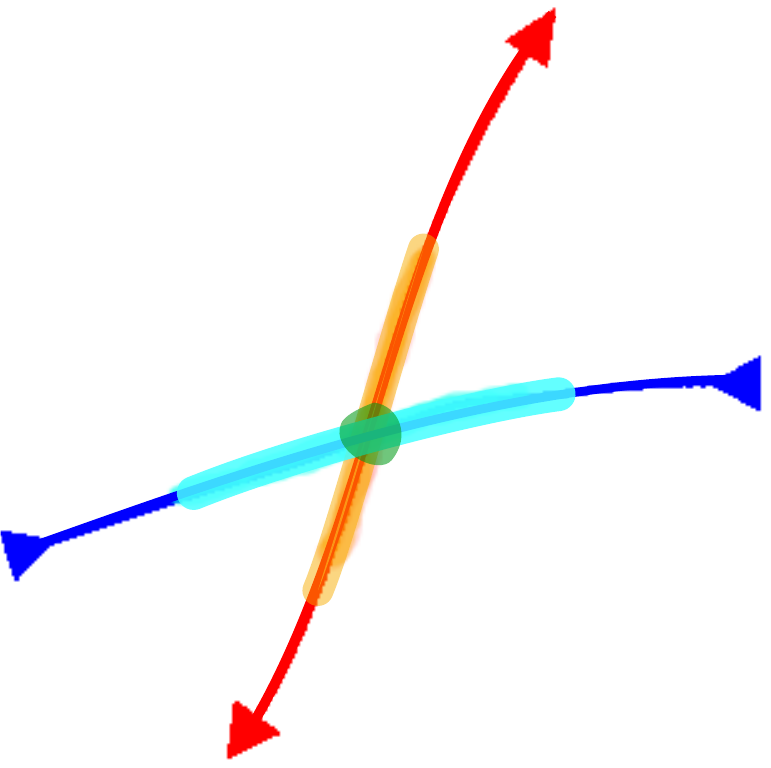} & \includegraphics[scale=0.35]{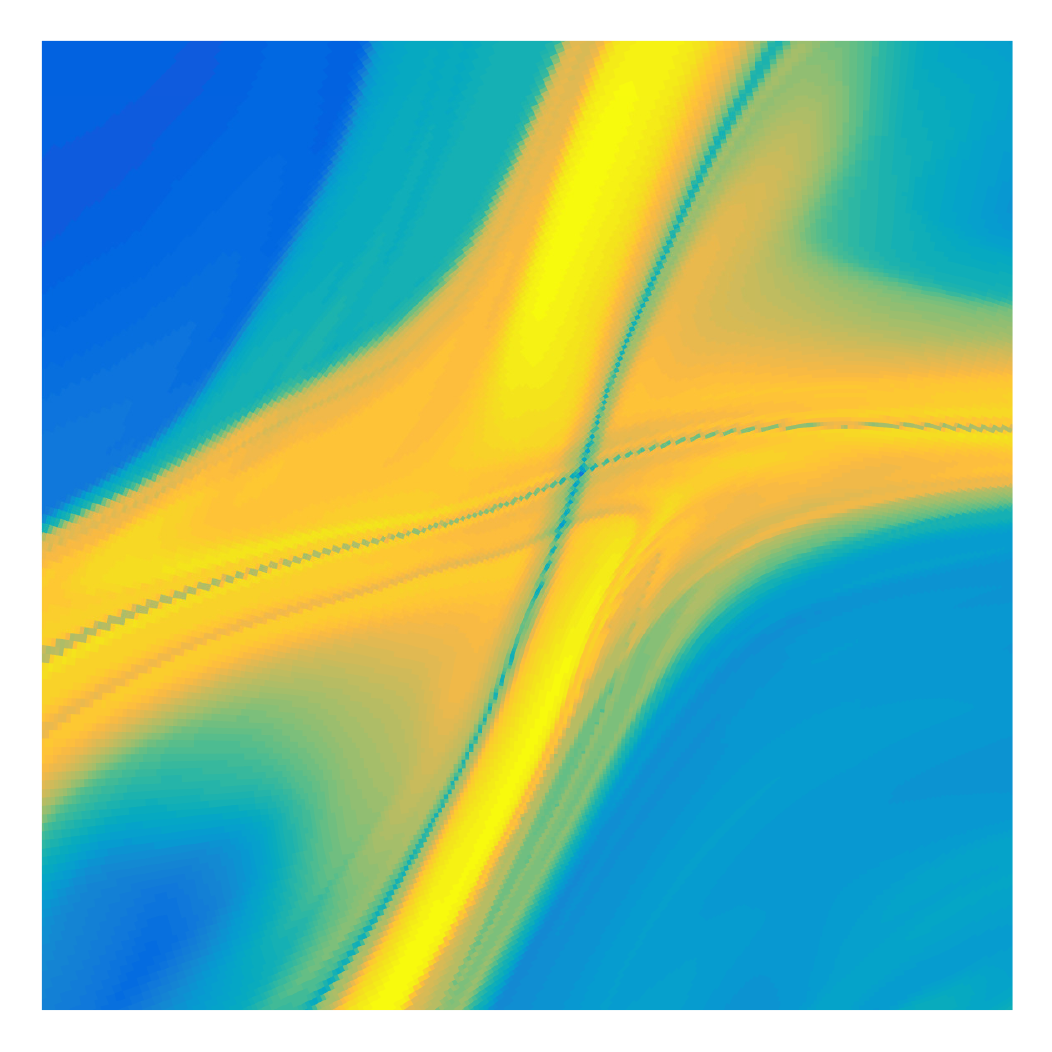}\\
\end{tabular}
 \caption{(a) Sketch of a hyperbolic  trajectory  and its stable (blue line) and unstable (red line) manifolds.  Particles in the green blob evolves forwards 
and backwards in time  
in time as the orange and cyan shades respectively. (b) the singular features of M are aligned with the stable and unstable manifolds. }\label{fig:HM}
\end{figure}

Trajectories of particles initially placed in the neighbourhood of a hyperbolic trajectory become aligned as time evolves forward with a curve called the {\em 
unstable manifold}; similarly,  trajectories of particles in the neighbourhood of a hyperbolic trajectory as time evolves backwards become aligned with a curve 
called the {\em stable manifold}. Figure \ref{fig:HM}(a) illustrate this forwards and backwards alignment of the green blobs with the orange and cyan blobs. 
For 
time dependent flows, hyperbolic trajectories do not correspond to hyperbolic instantaneous stagnation points of the velocity field. Their positions can be 
rather different, thus being the behavior of particles in a time interval quite counter-intuitive from what is observed for a velocity field that is frozen in 
time. 

Stable and unstable manifolds are aligned with singular features of $M$. Figure \ref{fig:HM}(b) shows two lines crossing in the middle of the yellowish region. 
These lines  are aligned with the stable and unstable manifolds appearing in panel (a) and correspond, respectively, to the stable and unstable manifolds of 
the 
hyperbolic trajectory at the crossing point.

Unstable and stable manifolds are {\em invariant curves}. This means that as time evolves either forwards or  backwards,  respectively, particles stay on those 
curves. Particles do not cross these curves, therefore they are material curves,  and are barriers to transport. Stable and unstable manifolds act as repelling 
and attracting material lines, respectively \cite{haller02}.

Apart from hyperbolic trajectories and their stable and unstable manifolds, other types of dynamical flow structures exist, in which particles tend to stay 
together, coherently, without dispersing, such as vortices that keep fluid parcels inside them, or jets. Invariant tori are the dynamical objects related to 
that behaviour.  These objects are invariant in the sense defined above, i.e. particles evolve in time staying on them. Further details on tori are given in 
Section 4.

Background for this Annex can be found in \cite{ottbook, warfm, physrep, samwig}.

 \section*{Acknowledgements}

J. Curbelo and A. M. Mancho are supported by MINECO grant MTM2014-56392-R. A. M. Mancho is supported by ONR. grant No. N00014-17-1-3003. C. R.
Mechoso was supported by the U.S. NSF grant AGS-1245069. The research of S. Wiggins is supported by ONR grant No. N00014-01-1-0769. {This paper has 
also received funding from the European Union's Horizon 2020 research and innovation programme
under the Marie Sklodowska-Curie grant agreement No 777822.}

 \bibliographystyle{unsrt}
 \bibliography{LD}
 
  \begin{tabular}{l}
\textbf{Jezabel Curbelo} \\
  {\small Departamento de Matem\'aticas} \\
 {\small Universidad Aut\'onoma de Madrid} \\
 {\small Instituto de Ciencias Matem\'aticas}\\
 {\small Campus de Cantoblanco UAM, 28049 Madrid, Spain.}\\
  {\small Email: jezabel.curbelo@uam.es} \\
\\

 \textbf{Carlos R. Mechoso}\\
  {\small            Department of Atmospheric and Oceanic Sciences,} \\
  {\small University of California at Los Angeles}\\
  \\
\textbf{Ana M. Mancho}\\ 
  {\small Instituto de Ciencias Matem\'aticas, CSIC-UAM-UC3M-UCM.}\\
  {\small Campus de Cantoblanco UAM, 28049 Madrid, Spain.}\\
    \\           
\textbf{Stephen Wiggins}\\
{\small School of Mathematics, University of Bristol.}\\
{\small Bristol BS8 1TW, UK.}

\end{tabular}

\end{document}